# Sputtered NbN Films for Ultrahigh Performance Superconducting Nanowire Single-Photon Detectors

Ilya A. Stepanov,[1] Aleksandr S. Baburin,[1, 2] Danil V. Kushnev,[1] Evgeniy V. Sergeev,[1, 2] Oksana I. Shmonina,[1] Aleksey R. Matanin,[1, 2] Vladimir V. Echeistov,[1] Ilya A. Ryzhikov,[1, 3] Yuri V. Panfilov,[1] and Ilya A. Rodionov[1, 2]

[1)]*FMN Laboratory, Bauman Moscow State Technical University, Moscow 105005, Russia*
[2)]*Dukhov Automatics Research Institute, (VNIIA), Moscow 127055, Russia*
[3)]*Institute for Theoretical and Applied Electromagnetics RAS, Moscow 125412, Russia*

(*Electronic mail: irodionov@bmstu.ru)

(Dated: 20 November 2023)

Nowadays ultrahigh performance superconducting nanowire single-photon detectors are the key elements in a variety of devices from biological research to quantum communications and computing. Accurate tuning of superconducting material properties is a powerful resource for fabricating single-photon detectors with a desired properties. Here, we report on the major theoretical relations between ultrathin niobium nitride (NbN) films properties and superconducting nanowire single-photon detectors characteristics, as well as ultrathin NbN films properties dependence on reactive magnetron sputtering recipes. Based on this study we formulate the exact requirements to ultrathin NbN films for ultrahigh performance superconducting nanowire single-photon detectors. Then, we experimentally study ultrathin NbN films properties (morphology, crystalline structure, critical temperature, sheet resistance) on silicon, sapphire, silicon dioxide and silicon nitride substrates sputtered with various recipes. We demonstrate ultrathin NbN films (obtained with more than 100 films deposition) with a wide range of critical temperature from 2.5 to 12.1 K and sheet resistance from 285 to 2000 Ω/sq, as well as investigate a sheet resistance evolution over for more than 40% within two years. Finally, we found out that one should use ultrathin NbN films with specific critical temperature near 9 K and sheet resistance of 400 Ω/sq for ultrahigh performance SNSPD.

## I. INTRODUCTION

State-of-the-art superconducting nanowire single-photon detectors (SNSPDs) show better performance compared to other single-photon detectors, including quantum efficiency up to 99%[1–3], counting rate higher than 1 GHz[4,5], jitter less than 3 ps[6], and dark count rate around $10^{-3}$ Hz[7]. The next innovations step makes it possible to create high-performance waveguide superconducting single-photon detectors[8] for low-loss photonic integrated circuits[9]. In combination with efficient single-photon sources[10,11] and low-loss integrated optical modulators[12] these are the key elements for an integrated photonic quantum computing platform[13]. Nevertheless, superconducting single-photon detectors which combines mentioned above characteristics all in one is still a difficult problem. Many recent studies demonstrate a strong dependence of SNSPD characteristics on nanowire material parameters[14–16]. However, a proper choice of NbN ultrathin film properties and its deposition recipe is still nonobvious everyday task for many scientific groups. In this paper, we report on a theoretical relations between SNSPD characteristics and niobium nitride (NbN) ultrathin films properties, as well as propose several approaches for getting the required SNSPD characteristics based on various films properties including high detection efficiency, high count rate and low dark count rate. Niobium nitride is one of the well-known material for high performance SNSPD as it has a high enough critical temperature $T_c$ allow using simply cryocoolers for operation. We experimentally investigate ultrathin NbN films deposited by reactive magnetron sputtering at different temperatures with various stoichiometries on silicon, sapphire, silicon dioxide and silicon nitride substrates. Then, a critical temperature $T_c$ and a sheet resistance $R_s$ of each deposited film were carefully measured. We investigated $T_c$ and $R_s$ dependencies on nitrogen concentration in the chamber and substrate temperature during deposition for different substrates. Also, it is well known that thin film properties could significantly change over time[17]. To study this effect, we routinely measured the $R_s$ degradation over two years due to NbN ultrathin film oxidation. More than 100 samples were investigated in the paper, providing a wide range of experimentally measured NbN ultrathin films properties.

## II. INFLUENCE OF MATERIAL PROPERTIES ON SNSPD PERFORMANCE

The most important parameters of superconducting films in term of future SNSPD applications are the critical temperature $T_c$ and sheet resistance at room temperature $R_s$. First, one chooses them for ultrathin NbN films characterization as they can be measured using an easy-to-implement four probe method. Second, they characterize superconducting films quite completely, reflecting the change in other film parameters, such as diffusion coefficient $D$, quasiparticle thermalization time $t_{th}$, etc. It is important to note that $T_c$ and $R_s$ are often related to each other in accordance with the universal scaling law[18], that means the lower the sheet resistance the higher the critical temperature. But in some cases, it is possible to get ultrathin NbN films with high $T_c$ and high $R_s$[19]. One of the key superconducting nanowire single-photon detector characteristic is a system detection efficiency (SDE) determining the probability that single photons which are sent to SNSPD will be detected. This parameter value is calculated by the follow-

ing equation:

$$SDE = \eta_{OCE} \cdot \eta_{ABS} \cdot \eta_{IDE} \quad (1)$$

where $\eta_{OCE}$ is an efficiency of optical coupling between an incident light and active area of SNSPD, which does not depend on film properties and is determined by a nanowire area[20], optical fiber and measuring setup parameters[1,21], $\eta_{ABS}$ is an absorption efficiency, which depends on a film thickness and its properties[22], nanowire fill factor[23], optical cavities[24], photon polarization[20] and wavelength[25], $\eta_{IDE}$ is an intrinsic detection efficiency, which depends on film properties, which directly determine detectors sensitivity to single photons[26], as well as a nanowire geometry[27], bias current and bath temperature[28], photon wavelength[29] and flux[1]. Another characteristic of SNSPD is a dark count rate (DCR), which describes a frequency of detector false counts. This undesirable effect can occur for a variety of reasons, including a current crowding effect[27], thermal fluctuations in superconductors[30], and phase slips[31] and depends on a bias current, bath temperature[29], external noise and photon flux[32]. DCR can be reduced by both selecting an optimal operating conditions for the SNSPD[23,26,32,33] and determining the most appropriate film properties[26] together with nanowire geometry[34–37].

In case of quantum communications application, for example, the most important SNSPD characteristics are count rate (CR) and jitter. Count rate is limited primarily by nanowire geometry[26] and is also depended on nanowire material[38], measurement setup[39], photon wavelength[40] and flux[1]. Jitter depends primarily on a cryogenic measurement setup and bias current[41], but is also determined by a nanowire geometry[42], bath temperature[43], external noise[44] and film properties[38]. However, the nature of jitter remains a poorly explored topic. Fig. 1 shows the impact of superconducting films properties, nanowire geometry, measurement setup and light properties on SNSPD characteristics. As one can see nanowire film properties influence significantly on the most important SNSPD characteristics. Here, we use some previously proposed approaches to evaluate the influence of ultrathin film properties on SNSPD absorption efficiency, counting rate and cut-off wavelength. We also propose a new model to link together critical temperature and sheet resistance of NbN films, as well as SNSPD operating parameters, with its dark counts due to thermal fluctuations.

### A. Absorption efficiency calculation

Achieving a high absorption efficiency is one of the biggest challenges in SNSPD development. One way to increase absorption is to form optical cavities under the nanowire, such as Distributed Bragg reflectors[45] and quarter-wave resonators[46]. Besides, nanowire film thickness influence greatly on absorption[45]. A high absorption efficiency is observed in thick superconducting films, but increased film thickness leads to decreased intrinsic efficiency of the SNSPD[47]; therefore, the film thickness usually does not exceed 10 nm. High absorption efficiencies without compromising other SNSPD characteristics can be achieved by optimizing superconducting film properties. Absorption of a thin metal film can be calculated as follows[48]:

$$\eta_{ABS} = \frac{4R_s/Z_0}{((R_s/Z_0)(n_{sub}+1)+1)^2} \quad (2)$$

where $Z_0 = 377\ \Omega$ is an impedance of free space, and $n_{sub}$ is substrate refractive index. To evaluate the absorption in a meander-shaped nanowire, the calculated value must be multiplied by a fill factor. The calculated absorption efficiency for NbN films depending on its sheet resistance and substrate refractive index is shown in Fig. 2(a), and the calculated $\eta_{ABS}$ for films on silicon, sapphire, silicon dioxide and silicon nitride substates are shown in Fig. 2(b). The substrates refractive indexes are taken for 1550 nm wavelength. It is worth noticing that in this chapter we are talking about nanowires fabricated on substrates without optical cavities. One can conclude that NbN film absorption can be increased either by using low refractive index substrates or with the certain sheet resistance at the extremum. The choice of a substrate material is often impossible due to technological route limitations for integrated devices. For example, SNSPD with self-alignment coupling[49] requires silicon substrate as it is fabricated with through Si etching[50]. Conversely, tuning the film sheet resistance can be done for every substrate and technology either by changing film thickness or deposition recipe. The figure shows an absorption extremum at the certain $R_s$, but it is extremely difficult to achieve such low resistance for sub-10nm NbN films. Summarizing the above, in order to achieve the highest absorption efficiency for ultrathin NbN films, its sheet resistance should be minimized.

### B. Count rate calculation

The count rate (CR) of the SNSPD is limited by the duration of the pulse resulting from photon counting. After the photon is absorbed, the nanowire transits from the superconducting state to normal one for time $\tau_{rise}$, followed by a voltage pulse and its decay for time $\tau_{fall}$. Thus, if the next photon arrives before the voltage pulse has completely decayed, it may not be detected, and the count rate of the SNSPD can be determined from the sum of the rise and fall times[51].

Since the rise and fall times depend on the kinetic inductance of the nanowire, which is determined by its geometry and material properties, the count rate can be calculated as[52]:

$$CR = \frac{1.764 k_B T_c \pi w (1-(\frac{T}{T_c})^4)}{\hbar R_s l} \left(\frac{R_L(R_N+R_L)}{2R_L+R_N}\right) \quad (3)$$

where $k_B$ is the Boltzmann constant, $w$ and $l$ are the width and length of the nanowire respectively, $T$ is the bath temperature, $\hbar$ is the reduced Planck constant, $R_N$ is the resistance of the nanowire that occurs when a photon is detected, and $R_L$ is the impedance of the coaxial electrical lines. $R_N$ is usually about 1 k$\Omega$[53], and $R_L$ is taken equal to 50 $\Omega$. For further calculations, the film thickness was taken equal to 5 nm, the width of the nanowire was 100 nm, and its length was 1 mm. The count rate calculation results for SNSPD based on NbN films with different properties are shown in Fig. 2(c).

| | | SYSTEM DETECTION EFFICIENCY | | | | |
|---|---|---|---|---|---|---|
| COUPLING EFFICIENCY | | ABSORPTION EFFICIENCY | INTRINSIC EFFICIENCY | DARK COUNT RATE | COUNT RATE | JITTER |
| · Distance between fiber end and SNSPD [1] · In-plane misalignment between optical fiber and SNSPD active area [21] · SNSPD active area [20] | Nanowire film properties | High [22] | High [26] | Low [26] | High [38] | Low [38] |
| | Nanowire geometry: | | | | | |
| | Thickness | High [22] | High [27] | No | High [26] | No |
| | Width | No | High [27] | No | High [26] | No |
| | Length | No | No | No | High [26] | Low [42] |
| | Fill-factor | High [23] | No | No | No | No |
| | Nanowire design | No | High [27] | High [27] | No | No |
| | Optical cavities | High [24] | No | No | No | No |
| | Bias current | No | High [28] | High [29] | High [40] | High [41] |
| | Bath temperature | No | High [28] | High [29] | No | Low [43] |
| | External noise | No | No | High [32] | No | High [44] |
| | Measurement setup | No | No | No | High [39] | Low [41] |
| | Light polarization | High [20] | No | No | No | No |
| | Photon wavelength | High [25] | High [29] | No | Low [40] | No |
| | Photon flux | No | High [1] | Low [32] | High [1] | No |

FIG. 1. SNSPD performance vs measurement setup, nanowire design, light and superconducting films properties. "High" indicates a strong impact of the parameter, "Low" - a weak impact, "No" - no information about the impact.

It can be seen that the count rate increases with an increase in the critical temperature and with a decrease in the sheet resistance of the film. The behavior of a superconductor, when a decrease in $T_c$ is associated with an increase in $R_s$, is typical and is well described by a universal scaling law[18]. In addition, a decrease in sheet resistance can also positively affect the absorption efficiency, as shown earlier. However, to achieve the best performance of the SNSPD, it is not preferable to take a film with the highest critical temperature and the smallest sheet resistance, since this could reduce the detection efficiency, what will be discussed below.

### C. Berezinskii–Kosterlitz–Thouless transition edge calculation

The most significant contribution to the total number of dark counts of the SNSPD is made by unbinding of vortices from pinning centers, as well as thermal fluctuations in a superconducting nanowire[54]. The unbinding of vortices from pinning centers due to an increase in the Lorentz force with increasing bias current can be prevented by optimizing the geometry of the SNSPD nanowire. Thermal fluctuations are the main source of dark counts at high bias currents and are associated, in particular, with the Berezinsky-Kosterlitz-Thouless (BKT) transition, which can occur at the temperatures above $T_{BKT}$[55–57]. Impact of thermal fluctuations on the characteristics on the SNSPD depends both on the bath temperature in the cryostat and on the properties of the nanowire thin film.

Based on the fact that $T_{BKT}$ can be determined from the sheet resistance of the film and its transition temperature $T_{c0}$, we took into account the effect of the bias current $I_{bias}$ on $T_{c0}$ and derived an equation to estimate $T_{BKT}$ for a superconducting film with critical temperature $T_c$ and sheet resistance $R_s$, biased by current $I_{bias}/I_c$ (the derivation is given in the supplementary):

$$T_{BKT} = \left(1 + 0.173 R_s \frac{e^2}{\hbar}\right)^{-1} \sqrt{1 - \frac{I_{bias}}{I_c}} T_c. \quad (4)$$

where $e$ is an elementary charge and $I_c$ is a critical current.

The results of $T_{BKT}$ calculation using Eq. (4) are shown in Fig. 2(d). The horizontal surface shows the bath temperature in the cryostat $T$. From the equation, it can be seen that the effect of sheet resistance on $T_{BKT}$ is negligible compared to the influence of $T_c$ and $I_{bias}$. Based on this, we present the surface plot at a constant sheet resistance of 500 Ω/sq. Fig. 2(e) shows the curve of intersection of the calculated surface and the horizontal planes corresponding to a widely occurring bath temperatures of 0.8, 1.2, 2.5 and 4.2 K.

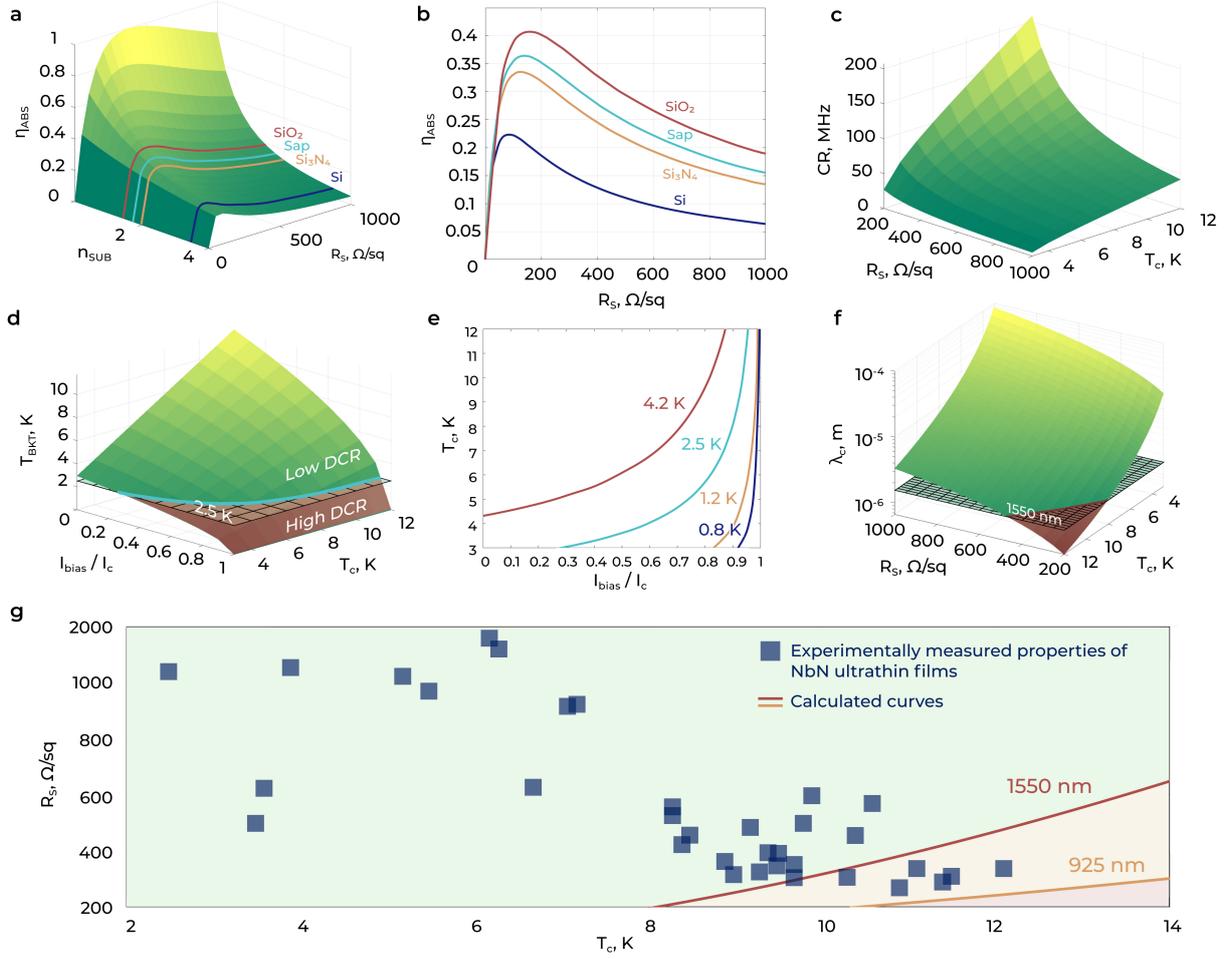

FIG. 2. Calculated dependencies of SNSPD characteristics on ultrathin NbN films properties. (a) Absorption efficiency for ultrathin NbN films with various sheet resistances on substrates with different refractive index ($n_{sub}$). (b) Absorption efficiency vs NbN film sheet resistance for silicon (Si), sapphire (Sap), silicon nitride ($Si_3N_4$) and silicon dioxide ($SiO_2$) substrates. (c) SNSPD count rate vs NbN films critical temperature and sheet resistance. (d) SNSPD Berezinsky-Kosterlitz-Thouless transition temperature vs NbN films critical temperatures and bias currents (with constant $R_s = 500\ \Omega/sq$). Horizontal surfaces correspond to the bath temperature. (e) Intersection curve of the calculated $T_{BKT}$ surface (Fig. 2(d)) and horizontal planes corresponded to 0.8, 1.2, 2.5 and 4.2 K. (f) SNSPD cut-off wavelength vs NbN films sheet resistance and critical temperature. (g) Intersection curve of the calculated $\lambda_c$ surface (Fig. 2(f)) and horizontal planes corresponded to 925nm and 1550nm. The blue dots show the experimentally measured properties of the sputtered ultrathin NbN films.

Thus, to ensure the minimum number of dark counts associated with thermal fluctuations, it is necessary to use ultrathin NbN films with a high critical temperature, which will weaken the influence of the bias current on the decrease in the energy gap. In addition, lowering the bath temperature can also greatly reduce the probability of thermal fluctuations occurrence at high bias currents. At this calculation stage, the requirements for the properties of the NbN film to minimize DCR do not contradict the requirements for increasing the count rate and absorption efficiency of the SNSPD.

### D. Cut-off wavelength calculation

The intrinsic detection efficiency $\eta_{IDE}$ of the SNSPD is determined both by the properties of the nanowire film and by the photon wavelength. As the wavelength increases, the photon energy decreases. At a certain wavelength $\lambda_c$ this energy becomes insufficient to effectively suppress superconductivity in a nanowire. This wavelength is called cut-off wavelength[27]. Thus, at wavelengths less than $\lambda_c$, unity intrinsic efficiency is observed, and at larger wavelengths, the efficiency is significantly reduced. When developing SNSPD, it is necessary to ensure that the cut-off wavelength is higher than the operating wavelength, which in our case is taken to be 1550 nm. This can be achieved by changing the geometry of the nanowire, reducing its width, but the limiting factors are the kinetic inductance, as shown above, and the technological capabilities of electron beam lithography. Further in this section, we aim to specify the properties of the NbN film to achieve unity intrinsic detection efficiency at wavelength of 1550 nm.

Based on the assumption that unity intrinsic detection efficiency is achieved when the order parameter in the hot spot is fully suppressed the cut-off wavelength could be calculated as follows[52]:

$$\lambda_c = \frac{hcR_s e^2}{\pi \tau_{th}(k_B T_c)^2 \left( \frac{1.764^2 \left(1-\frac{T}{T_c}\right)}{2} + \frac{2\pi^2}{3\left(1+\frac{C_{ph}(T_c)}{C_e(T_c)}\right)} \right)} \quad (5)$$

where $h$ is a Planck constant, $t_{th}$ is thermalization time of quasiparticles, $C_{ph}$ and $C_e$ are the phonon and electron heat capacities of the film, respectively, and $T$ is bath temperature. For further calculation, the thermalization time was taken equal to 7 ps, and the ratio of phonon and electron heat capacities was taken equal to 4.08[58]. Assuming a bath temperature of 2.5 K, we plotted the dependence of the cut-off wavelength on the properties of the NbN film shown in Fig. 2(f). For the cut-off wavelengths equal to 925 and 1550 nm, we obtain the curves shown in Figure 2(g), where films located above the curve will provide unity intrinsic detection efficiency at the considered wavelength, while films located below do not. Looking ahead, the dots show the properties of our ultrathin NbN films, the deposition of which is discussed later in this paper.

To ensure a guaranteed high cut-off wavelength, it is possible to use a film with a low critical temperature and high sheet resistance, but this contradicts the requirements presented earlier for achieving low DCR, high CR and high $\eta_{ABS}$. In order to obtain a specific ratio of SNSPD characteristics, one can use the given dependencies and achieve an improvement in one characteristic at the expense of a deterioration in another. However, in order to achieve complex high characteristics of SNSPD, we propose the following approach. To ensure unity intrinsic detection efficiency and at the same time to obtain high absorption, high count rate and low dark count rate, it is necessary to use NbN films corresponding to the points in Fig. 2(g) located above the curve, but as close to it as possible. The possibility NbN films with specific properties deposition is necessary to create SNSPDs with predetermined characteristics. The remainder of this paper is devoted specifically to recipes of NbN films deposition in a wide range of properties to enable the fabrication of SNSPDs with a wide range of characteristic ratios.

## III. STUDY OF ULTRATHIN NBN FILMS

### A. Magnetron sputtering system

All thin films were deposited by ultra-high vacuum reactive magnetron sputtering. The schematic layout of the sputtering system vacuum chamber is shown in Fig. 3(a). The vacuum chamber is pumped out by a cryogenic pump, which ensure the base ultra-high vacuum conditions before deposition processes. In all the experiments described below the sputtering current, argon flow, sputtering angle, distance between samples and magnetron source, and operating pressure remained constant. The deposition angle was chosen to ensure the best uniformity of film thickness over the substrates[59]. The nitrogen flow was controlled by a mass flow controller and varied to obtain NbN films with different stoichiometry. The substrate temperature was varied to obtain films with different crystalline structures. It was set using located above the sample holder infrared heaterspower, and controlled using a temperature sensor located nearby. More detailed information on the deposition and characterization methods for NbN films can be found in the supplementary materials.

### B. Ultrathin NbN films sputtering and characterization

We chose four types of substrates: high-resistivity silicon (10000 Ω), sapphire, silicon dioxide on silicon, and silicon nitride on silicon. We varied the nitrogen flow to achieve its concentration in the vacuum chamber during deposition from 10 to 40 percent, the substrate temperature was varied from 21 to 800 °C. 5nm-thick NbN films were chosen for SNSPD fabrication. However, it is very difficult in some cases to use quartz crystal thickness control for accurate ultrathin film deposition as deposition rate also depends on the substrate temperature (quartz crystal cannot be heated and controlled correctly at high deposition temperatures). Therefore, to deposit 5nm-thick NbN films with various recipes, we first deposited the films with the thickness from 50 to 100 nm using the same deposition parameters, hereinafter referred to as thin films. Firstly, it allowed us to accurately measure the film thickness using scanning electron microscope (SEM) and carefully calculate deposition time for 5nm-thick films. Secondly, the thicker films allowed us to precisely characterize the films surface and particular feature of films growth using SEM. SEM images of the 50nm-thick NbN films deposited in characteristic regimes on silicon and sapphire substrates are shown in Fig. 3(b) and (c), respectively. For films deposited on silicon dioxide and silicon nitride substrates, the surface is very similar to that shown for silicon, so these images are provided in the supplementary material.

One can conclude that 50nm-thick films surface on the silicon substrate obtained at room temperature and at 400 °C look almost the same at any nitrogen concentration. The surface is represented by multiple small grains, which do not have a specific direction of growth or clustering. However, the films morphology sputtered at 800 °C differs dramatically. The film deposited at 10% nitrogen concentration has a transitional structure with multiple individual grains of much larger size. Films obtained at 20% and 35% nitrogen concentrations consist of individual grain clusters with characteristic dimensions of tens nanometers. Here, we demonstrated together with the well-known effect of grain sizes increasing at elevated substrate temperature[60], that nitrogen promotes the larger grain clusters formation. For the films on sapphire substrates the effect of the nitrogen concentration on the morphology is noticeable even at low temperatures. At room temperature, with the increased nitrogen concentration the larger grain size is observed, which become elongated, but does not have a clear direction of growth. By increasing the nitrogen concentration at 400 °C one can observe the same character of morphology modification. However, at higher nitrogen concentrations the grains acquire several characteristic growth directions, which are mixed over the substrate area. At 800 °C the growth behavior is almost the same as for silicon substrates, while nitrogen also promotes a larger clusters formation, but on sapphire substrates they have several micrometers scale dimensions. We observed elongated grains with a strict direction direction inside these clusters.

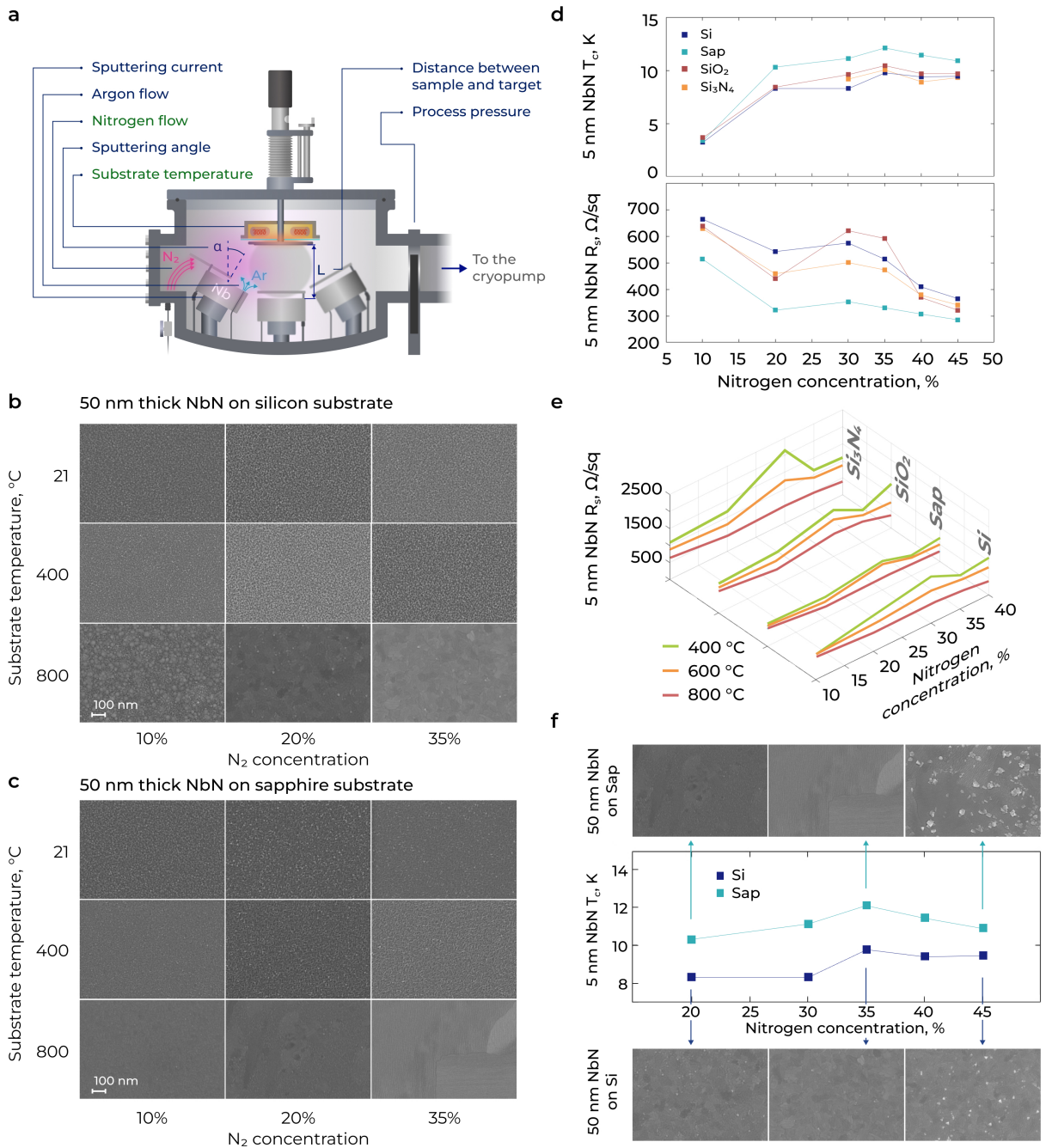

FIG. 3. Experimentally measured properties of sputtered NbN thin films. (a) Vacuum chamber of the magnetron sputtering system. Fixed parameters are shown in navy and variable parameters are shown in green. (b) SEM images of 50nm-thick NbN films surfaces on silicon substrates with various substrate temperature and nitrogen concentration during sputtering. (c) SEM images of 50nm-thick NbN films surfaces on sapphire substrates with various substrate temperature and nitrogen concentration during sputtering. (d) Critical temperature and sheet resistance of 5nm-thick NbN films on various substrates depending on the nitrogen concentration (sputtered at 800 °C substrate temperature). (e) 5nm-thick NbN films sheet resistance vs nitrogen concentration for various substrates and deposition temperatures. (f) 5nm-thick NbN films critical temperatures on silicon and sapphire substrates vs nitrogen concentration (sputtered at 800 °C substrate temperature). The relations between 5nm-thick films and 50nm-thick films sputtered with the same recipes critical temperatures and morphology are demonstrated.

Based on these data we sputtered 5nm-thick NbN films on the same substrates at room temperatures, 400, 600 and 800 °C with various nitrogen concentrations, hereinafter referred to as ultrathin films. The deposition of ultrathin NbN films at room temperature did not allow us to obtain sheet resistances less than 1000 Ω/sq and high critical temperatures, therefore, we did not consider these films for further SNSPD fabrication. With an elevated temperature up to 800 °C we obtained films with higher critical temperatures and lower sheet resistances. The dependence of critical temperature and sheet resistance on nitrogen concentration for ultrathin NbN films deposited at 800 °C is shown in Fig. 3(d). One can notice that at 35% nitrogen concentration an extremum of the critical temperature is observed, while the sheet resistance tends to decrease with nitrogen concentration increasing. However, we found that at 30% nitrogen concentration the sheet resistance shows a slight local increase. For the films deposited at lower temperature the local increase in $R_s$ was much more significant, as can be seen in Fig. 3(e). At all the considered substrate temperatures and nitrogen concentrations ultrathin NbN films on the sapphire substrates showed the highest critical temperatures and the lowest sheet resistances, while the films on silicon, on the contrary, tended to have the lowest critical temperatures and the highest sheet resistances. The ultrathin NbN films on the silicon nitride and silicon dioxide substrates properties values were close to each other and were found between the values for the films on the silicon and sapphire substrates. We deposited ultrathin NbN films on various substrates at 800 °C temperature and 30-40% nitrogen concentration, which show properties that can theoretically ensure SNSPD with high performance (high detection efficiency, high count rate and low dark count rate) as it is presented earlier in this paper.

Another interesting thing is a relationship between the film morphology and its superconducting properties, since it is widely known that the properties of a film are largely determined by its structure[61]. SEM method does not allow us to evaluate ultrathin NbN films morphology; however, indirectly, the trends in morphology variation with a different deposition recipes can be assessed on thicker films. Fig. 3(f) demonstrates the correspondence of deposited at the same parameters 5nm-thick films and 50nm-thick films critical temperatures and morphology of. We found out that with increased nitrogen concentration (up to a certain value) both the grain size of 50nm-thick film and the critical temperature of 5nm-thick film increase. When 50nm-thick films reach the maximum grain size (at 35% nitrogen concentration), the extremum of $T_c$ is observed for 5nm-thick films. With the following increase in nitrogen concentration from 35 to 45% the grain clusters size does not grow, however, multiple crystallites appear that protrude above the film surface. Thus, we found out that the maximum critical temperature of 5nm-thick films corresponds to 50nm-thick films morphology with the largest grain clusters, but multiple inclusions with individual out-of-plane growth directions are not observed for them.

It is critical for high performance SNSPD fabrication, that ultrathin NbN films change their properties over time due to oxidation[62]. We investigated the sheet resistance of our 5nm-thick NbN films over time and found that $R_s$ could increase by more than by more than 40% within two years and the rate of its change with time is decreased; the dependence of $R_s$ increase over time has a power-law form. We confirmed that the sheet resistance increase is mainly triggered by interaction of the films with the atmosphere (for a detailed description see the supplementary).

## IV. DISCUSSION

In conclusion, based on the theoretical model we demonstrated that SNSPD low dark count rate, high count rate and high absorption can be achieved by reducing NbN nanowire ultrathin film sheet resistance and increasing its critical temperature, however, it contradicts achieving the high intrinsic detection efficiency. We found out that in order to create high performance SNSPD (high intrinsic detection efficiency, high absorption efficiency, high count rate and low dark count rate), one should use ultrathin NbN films with specific properties as follows: the critical temperature near 9 K and sheet resistance of 400 Ω/sq. We demonstrated the films with the desired properties experimentally in this paper. We investigated several batches of NbN thin films deposited by reactive magnetron sputtering on Si, $Al_2O_3$, $SiO_2$ and SiN substrates at different temperatures and nitrogen concentrations. For all the sputtered films on different substrates we noticed the following trends: 1) as the substrate temperature increases, the critical temperature increases, but the sheet resistance decreases; 2) by increasing nitrogen concentration, $T_c$ shows a parabolic dependence with an extremum at a certain nitrogen concentration, and the sheet resistance tends to decrease. We concluded that in order to deposit 5nm-thick NbN films with the highest possible critical temperature and low sheet resistance, that is, to achieve a high intrinsic detection efficiency, one should use the high substrate temperature and certain nitrogen concentration (35% for our deposition tool). On the contrary, to get low $T_c$ and high $R_s$, regardless of the nitrogen concentration the substrate must be cooled down to room temperature. Finally, we experimentally deposited and analyzed superconducting NbN thin films with a wide range of properties, including $T_c$ from 2.5 to 12.1 K and $R_s$ from 285 to more than 2000 Ω/sq. We investigated the dependence of ultrathin NbN films sheet resistance over time and demonstrated that it could increase by more than 40% within two years for some samples.

## SUPPLEMENTARY MATERIAL

Additional data and characterization relevant to this article and referenced in the main text are provided in the supplementary material.

## ACKNOWLEDGMENTS

Samples were fabricated and studied at the BMSTU Nanofabrication Facility (FMN Laboratory, FMNS REC, ID 74300).

## AUTHOR DECLARATIONS

### A. Conflict of Interest

The authors declare no competing interests.

### B. Author Contributions

**Ilya A. Stepanov**: Conceptualization (lead); Investigation (lead); Formal analysis (equal); Methodology (lead); Project administration (equal); Visualization (equal); Writing – original draft (lead). **Aleksandr S. Baburin**: Conceptualization (equal); Formal analysis (equal); Project administration (equal); Writing – original draft (equal); Writing – review & editing (equal). **Danil V. Kushnev**: Formal analysis (supporting); Investigation (supporting); Methodology (equal). **Evgeniy V. Sergeev**: Investigation (supporting); Methodology (equal). **Oksana I. Shmonina**: Visualization (equal); Formal analysis (supporting). **Aleksey R. Matanin**: Investigation (equal); Methodology (supporting). **Vladimir V. Echeistov**: Investigation (supporting); Methodology (equal). **Ilya A. Ryzhikov**: Conceptualization (equal); Formal analysis (equal); Writing – review & editing (equal). **Yuri V. Panfilov**: Conceptualization (equal); Formal analysis (equal). **Ilya A. Rodionov**: Conceptualization (equal); Formal analysis (equal); Project administration (lead); Supervision (lead); Writing – original draft (equal); Writing – review & editing (lead).

## DATA AVAILABILITY STATEMENT

The data that support the findings of this study are available within the article and its supplementary material and from the corresponding author upon reasonable request.

# Sputtered NbN Films for Ultrahigh Performance Superconducting Nanowire Single-Photon Detectors

Ilya A. Stepanov[1], Aleksandr S. Baburin[1,2], Danil V. Kushnev[1], Evgeniy V. Sergeev[1,2], Oksana I. Shmonina[1], Aleksey R. Matanin[1,2], Vladimir V. Echeistov[1], Ilya A. Ryzhikov[1,3], Yuri V. Panfilov[1], Ilya A. Rodionov[1,2,*]

[1]FMN Laboratory, Bauman Moscow State Technical University, Moscow, Russia

[2]Dukhov Automatics Research Institute, (VNIIA), Moscow, Russia

[3]Institute for Theoretical and Applied Electromagnetics RAS, Moscow, Russia

*e-mail: irodionov@bmstu.ru

This supplement provides equations and experimental details to support the claims made in the main text. First, we present the derivation of the equation for calculating the Berezinskii–Kosterlitz–Thouless temperature based on the properties of the film. We then show high-resolution SEM images of the surfaces of NbN thin films on silicon, sapphire, silicon dioxide, and silicon nitride substrates. Next, we describe a study of the degradation of ultrathin NbN films over time and analyze the reasons for the increase in its sheet resistance. Finally, we provide information on the research methods, including reactive magnetron sputtering, sheet resistance and critical temperature measurements, and scanning electron microscopy.

### 1. Berezinskii–Kosterlitz–Thouless temperature calculation

The Berezinskii–Kosterlitz–Thouless transition temperature $T_{BKT}$, above which thermal fluctuations make a significant contribution to the dark count rate of the detector, can be calculated as [1]:

$$T_{BKT} = \left(1 + 0.173 R_s \frac{e^2}{\hbar}\right)^{-1} T_{c0}, \quad (S1)$$

where $R_s$ is the sheet resistance of the film, $e$ is an elementary charge, $\hbar$ is the reduced Planck constant, and $T_{c0}$ is the transition temperature of the film.

The above equation shows that $T_{BKT}$ of a thin film depends only on its sheet resistance $R_s$ and the transition temperature $T_{c0}$. For a nanowire to which a bias current $I_{bias}$ is applied, it must be taken into account that $T_{c0}$ strongly depends on the $I_{bias}$. Thus, it is correct to consider the transition temperature of the film measured at negligibly small bias current the critical temperature $T_c$. When a large bias current flows through a superconductor, especially close to the critical current $I_c$, the transition temperature decreases in accordance with the equation [2]:

$$\frac{T_{c0}}{T_c} = \sqrt{1 - \frac{I_{bias}}{I_c}}. \quad (S2)$$

The need to apply a bias current close to $I_c$ in order to achieve high detection efficiency can significantly reduce the transition temperature, therefore, at some bias current, $T_{BKT}$ becomes lower than the bath temperature $T$, which will correspond to a phase with no long-range order and lead to a large number of dark counts of SNSPD due to thermal fluctuations. By combining equations (S1) and (S2), it is possible to estimate $T_{BKT}$ for a superconducting film with critical temperature $T_c$ and sheet resistance $R_s$, biased by current $I_{bias}/I_c$:

$$T_{BKT} = \left(1 + 0.173 R_s \frac{e^2}{\hbar}\right)^{-1} \sqrt{1 - \frac{I_{bias}}{I_c}} T_c. \tag{S3}$$

2. **SEM images of the surfaces of NbN thin films**
 2.1. **NbN thin films on silicon substrate**

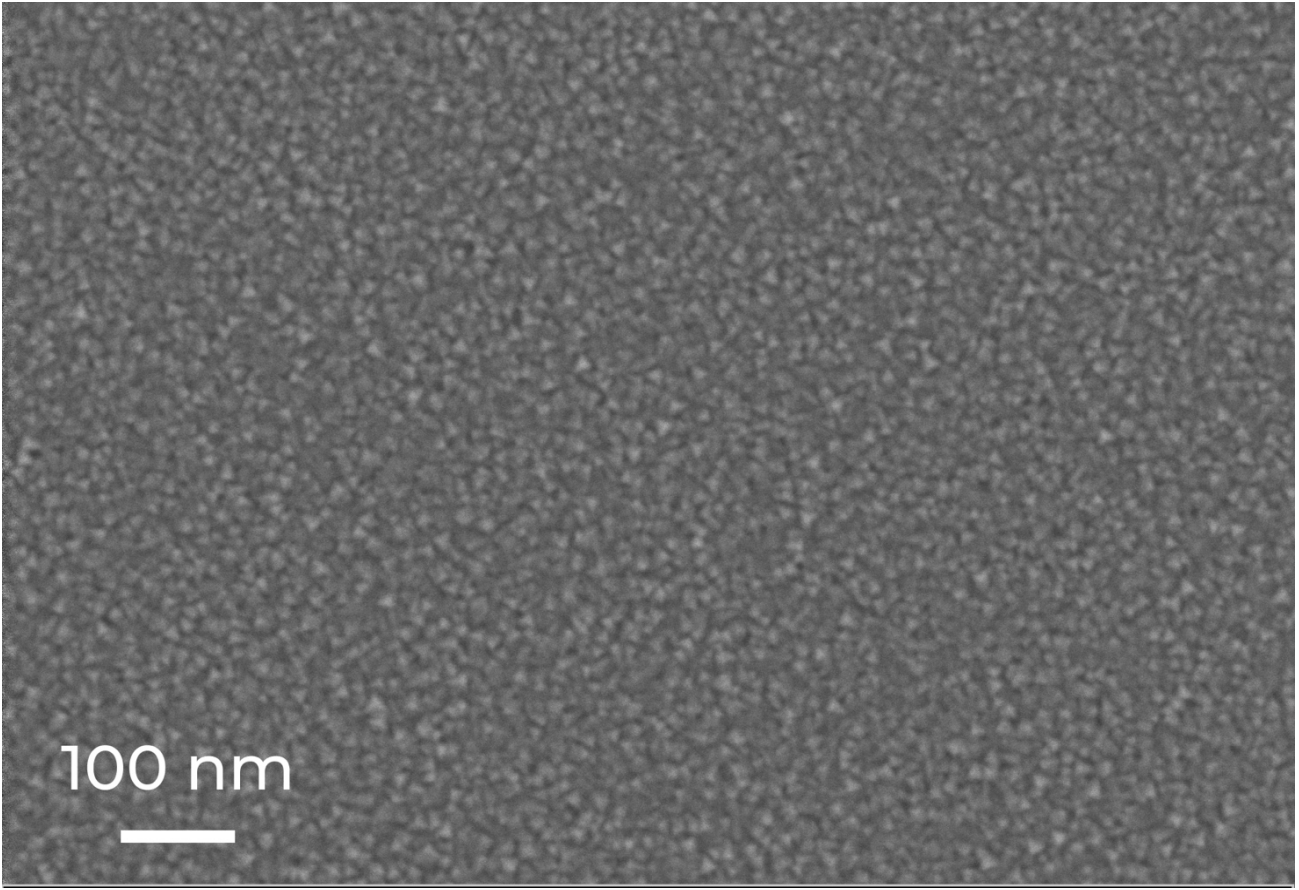

**Figure S1.** SEM image of the surface of an NbN thin film deposited at a temperature of 21 °C and a nitrogen concentration of 10% on a silicon substrate

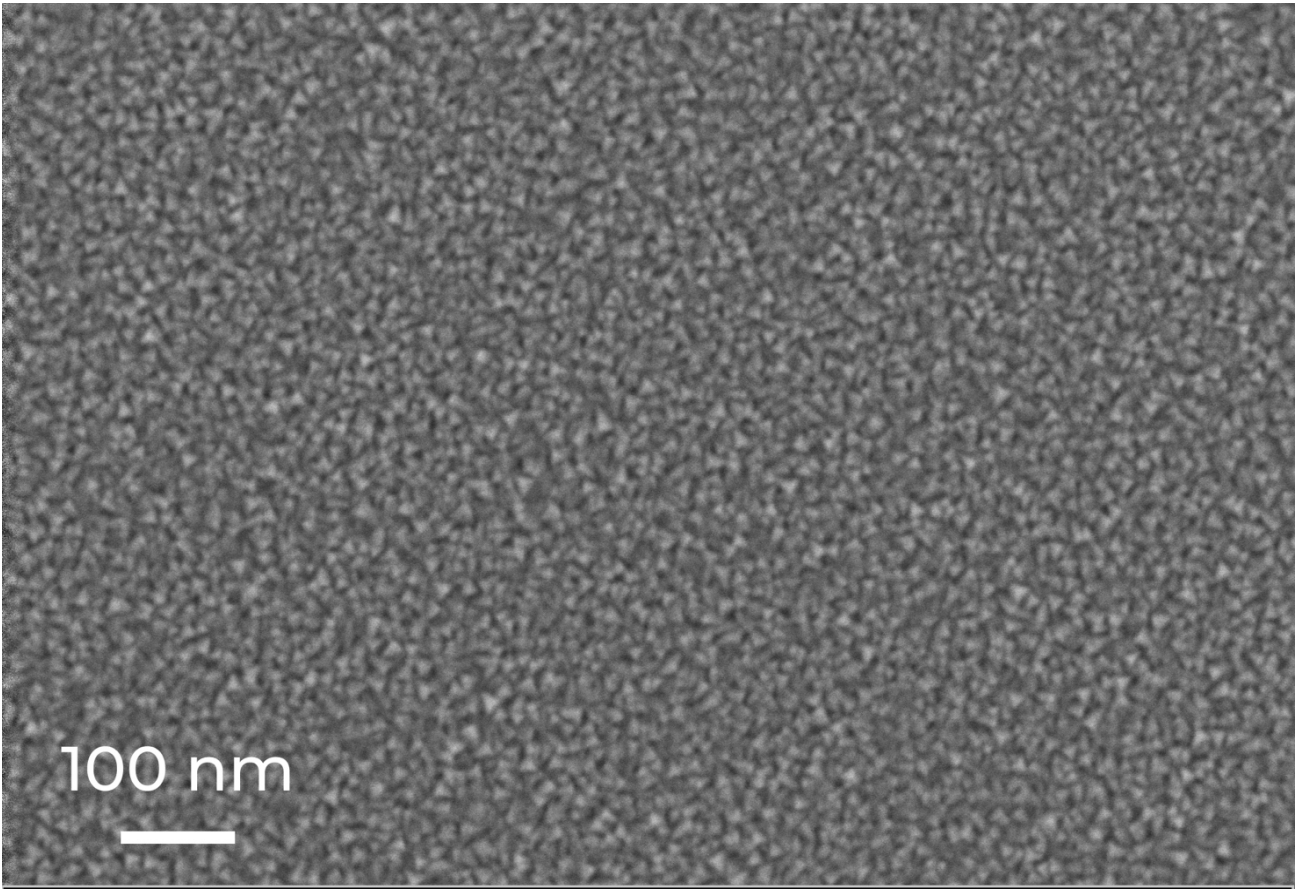

**Figure S2.** SEM image of the surface of an NbN thin film deposited at a temperature of 21 °C and a nitrogen concentration of 20% on a silicon substrate

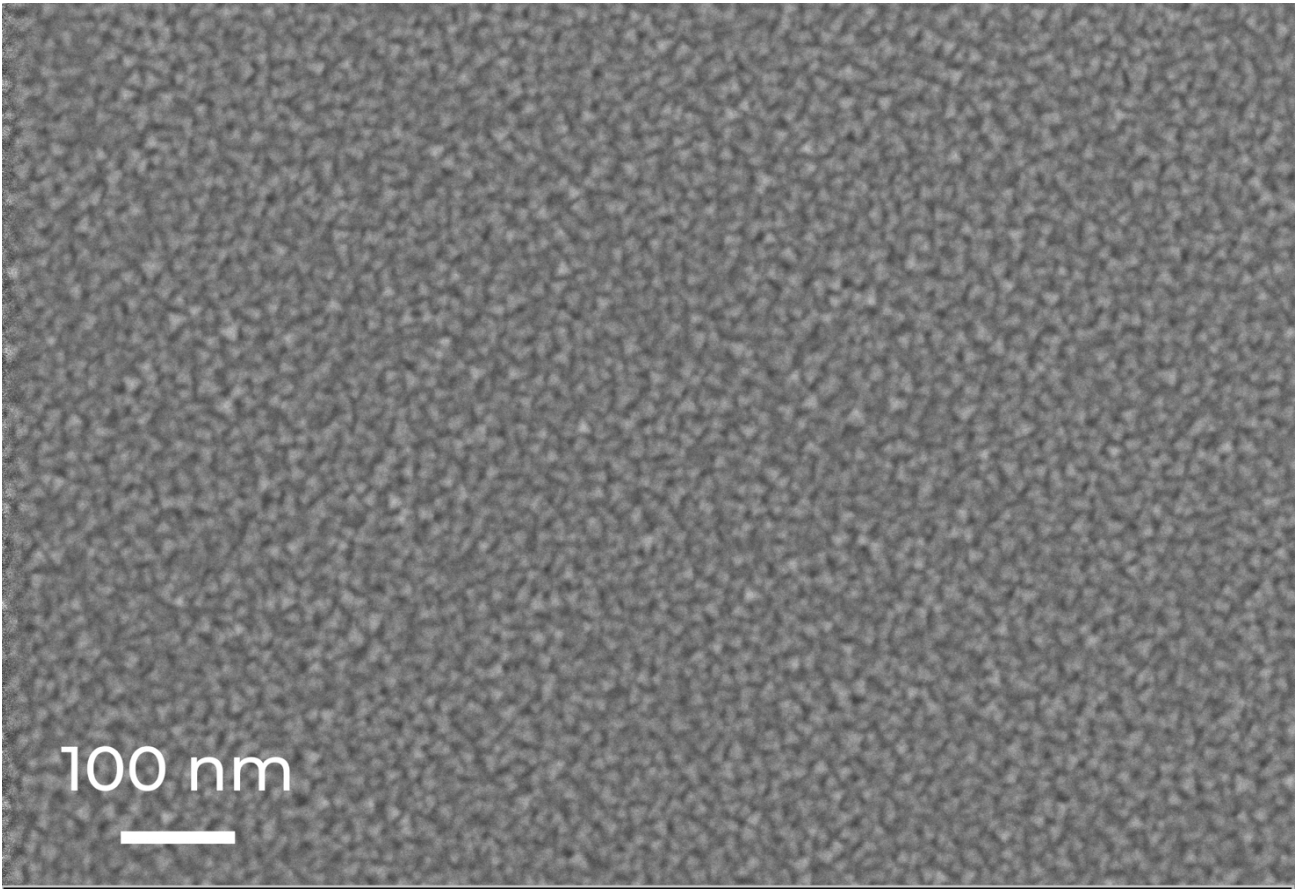

**Figure S3.** SEM image of the surface of an NbN thin film deposited at a temperature of 21 °C and a nitrogen concentration of 35% on a silicon substrate

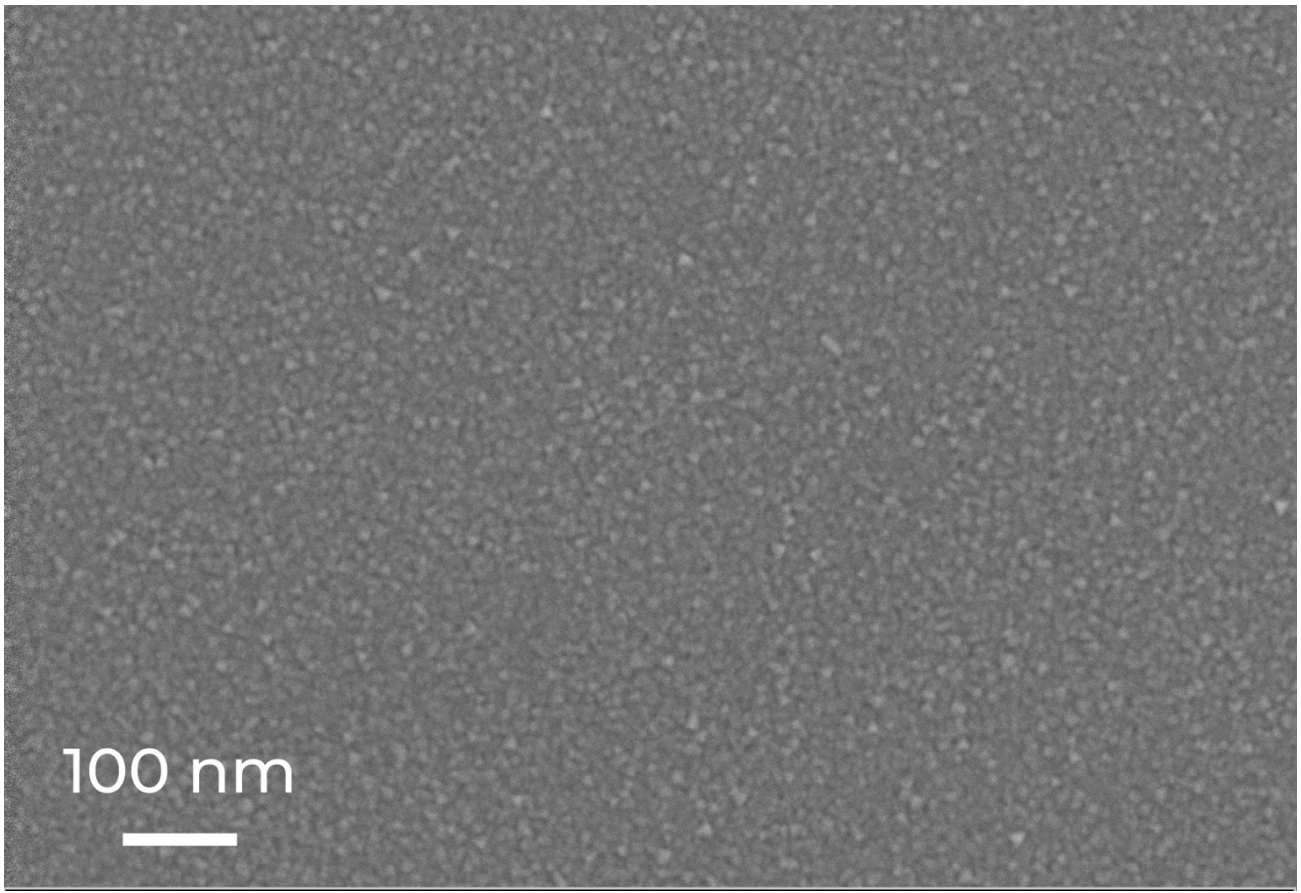

**Figure S4.** SEM image of the surface of an NbN thin film deposited at a temperature of 400 °C and a nitrogen concentration of 10% on a silicon substrate

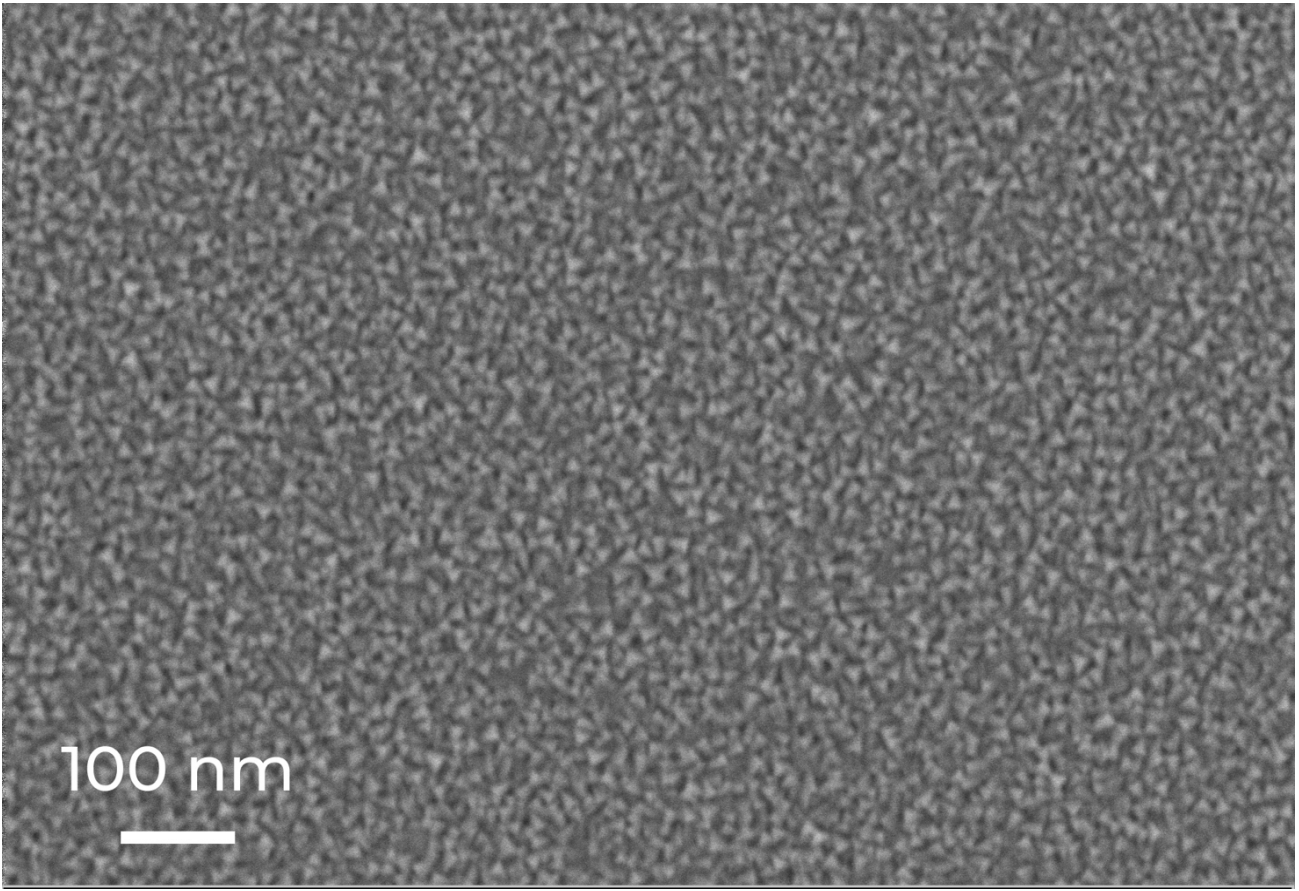

**Figure S5.** SEM image of the surface of an NbN thin film deposited at a temperature of 400 °C and a nitrogen concentration of 20% on a silicon substrate

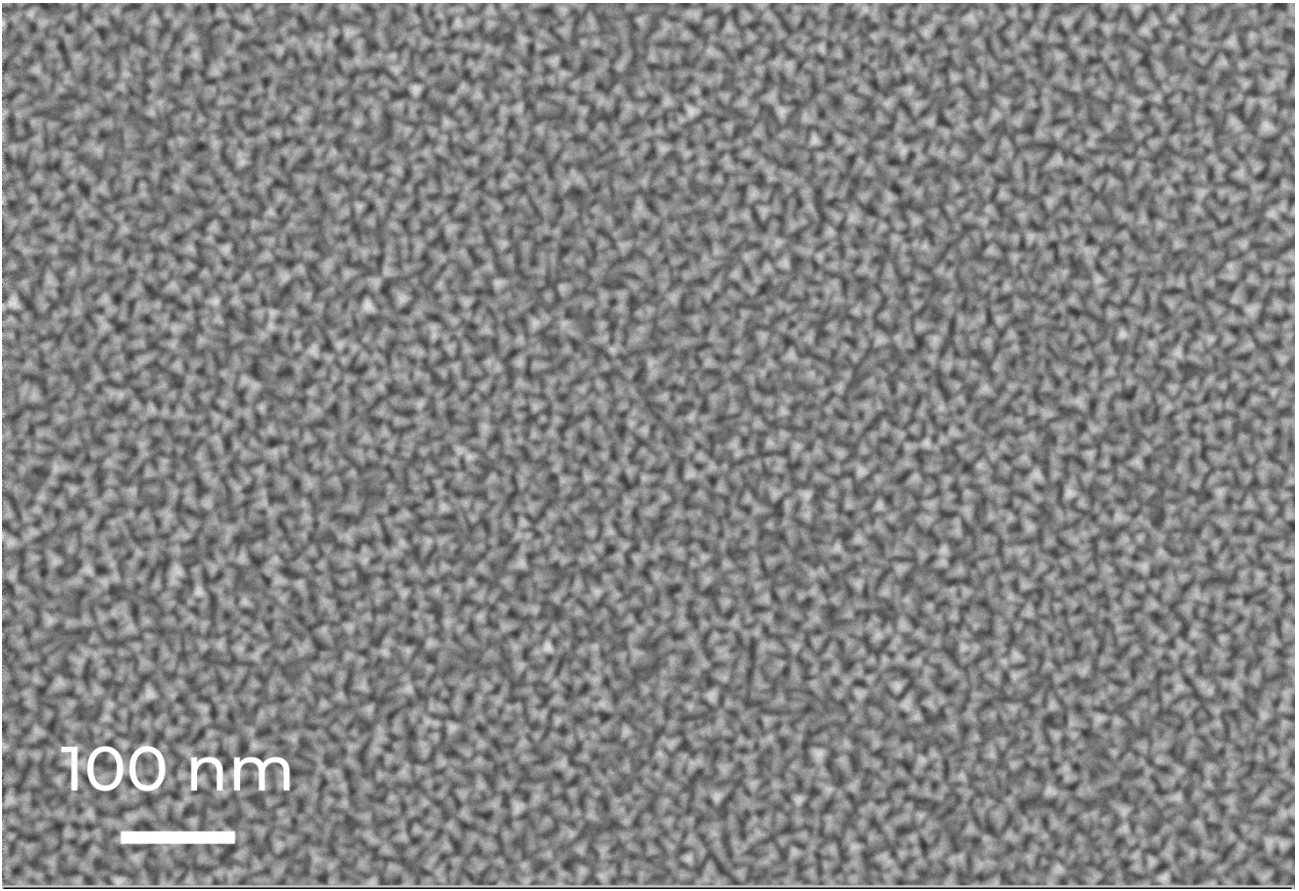

**Figure S6.** SEM image of the surface of an NbN thin film deposited at a temperature of 400 °C and a nitrogen concentration of 35% on a silicon substrate

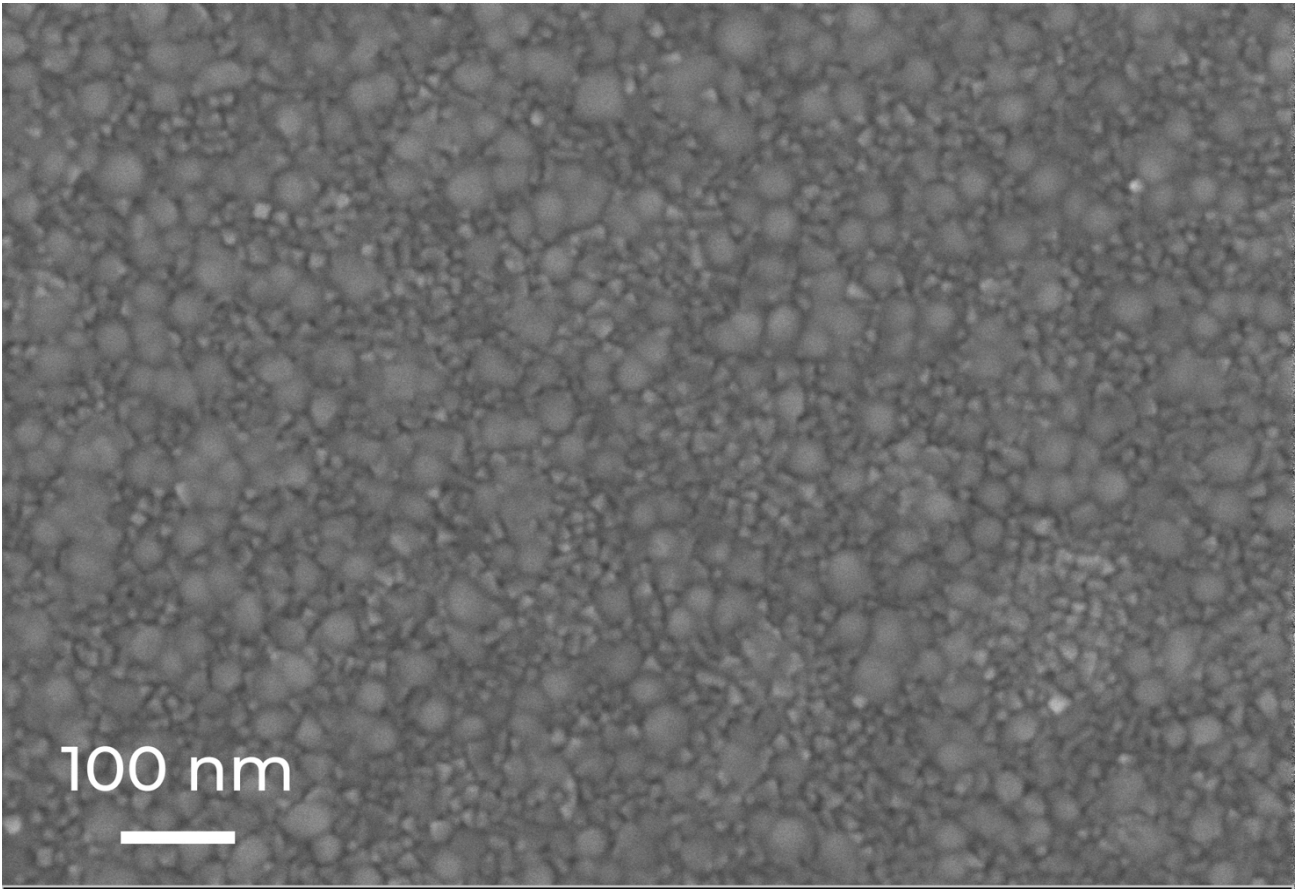

**Figure S7.** SEM image of the surface of an NbN thin film deposited at a temperature of 800 °C and a nitrogen concentration of 10% on a silicon substrate

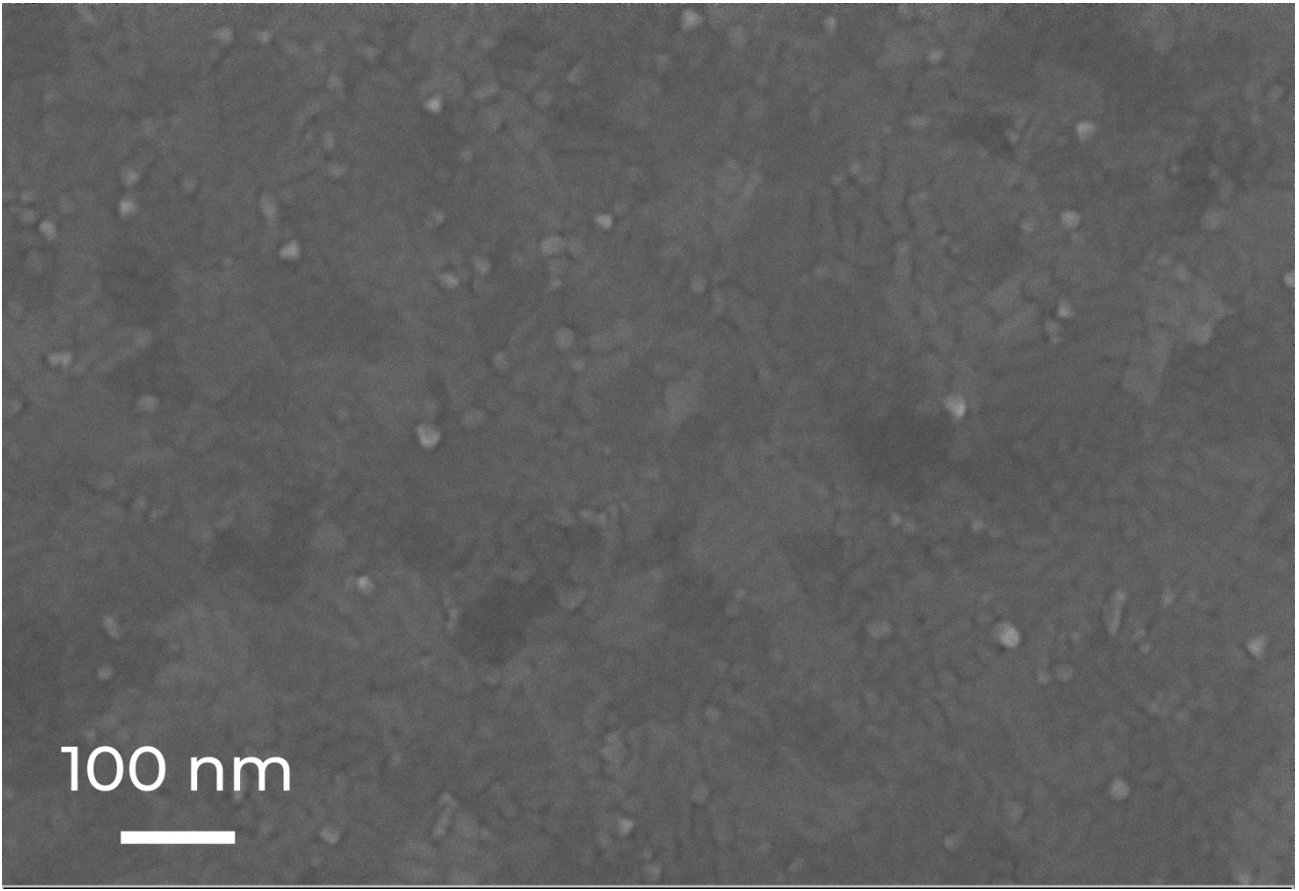

**Figure S8.** SEM image of the surface of an NbN thin film deposited at a temperature of 800 °C and a nitrogen concentration of 20% on a silicon substrate

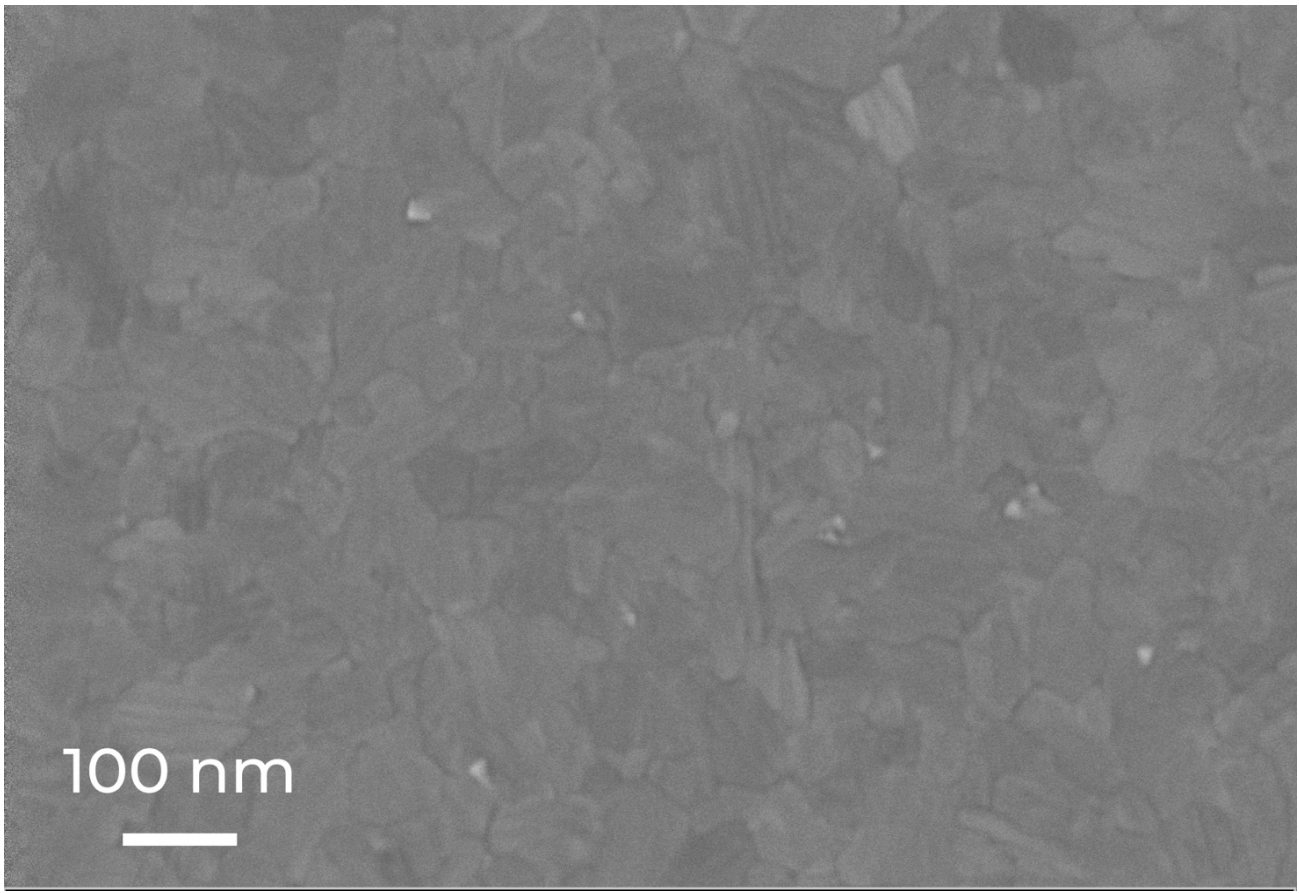

**Figure S9.** SEM image of the surface of an NbN thin film deposited at a temperature of 800 °C and a nitrogen concentration of 35% on a silicon substrate

**2.2. NbN thin films on sapphire substrate**

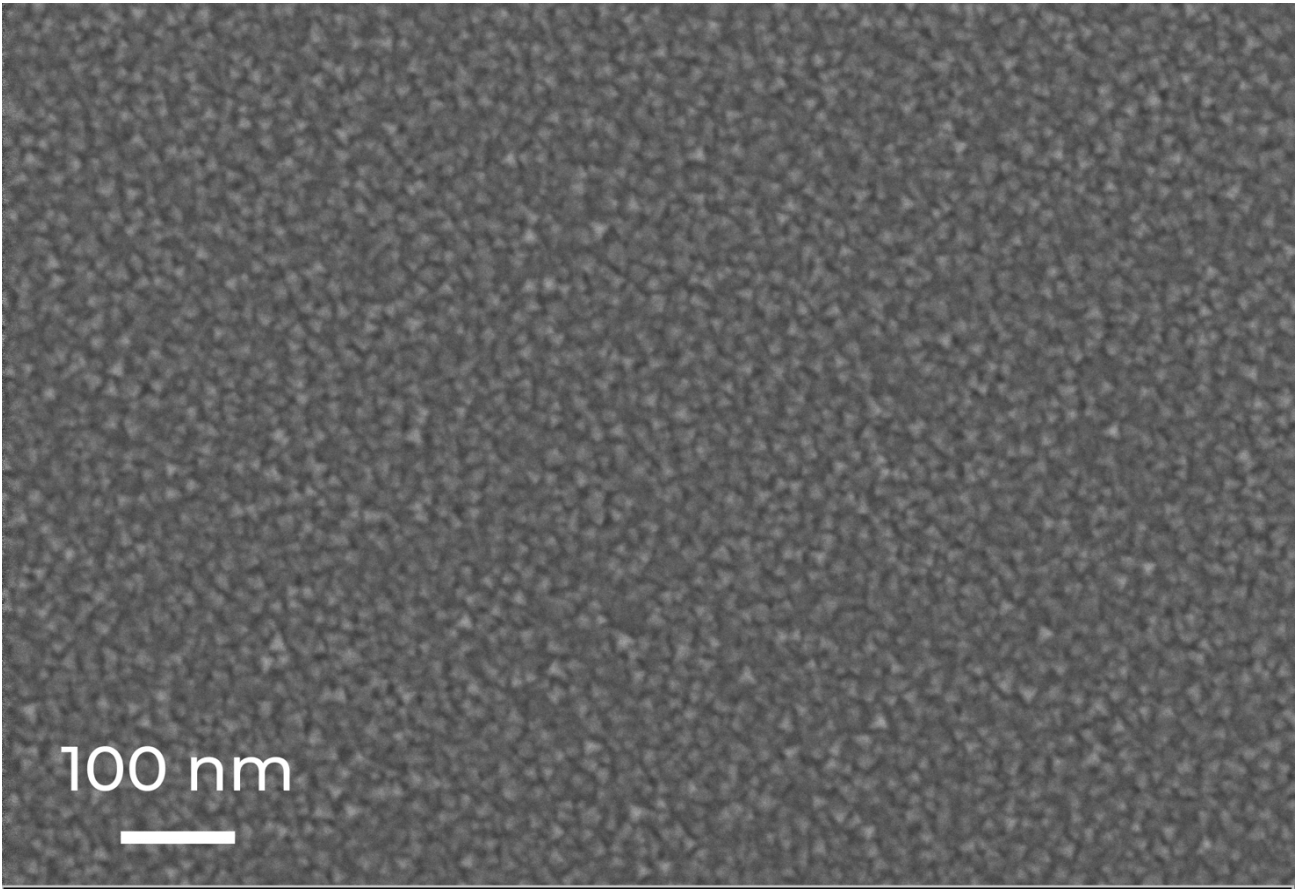

**Figure S10.** SEM image of the surface of an NbN thin film deposited at a temperature of 21 °C and a nitrogen concentration of 10% on a sapphire substrate

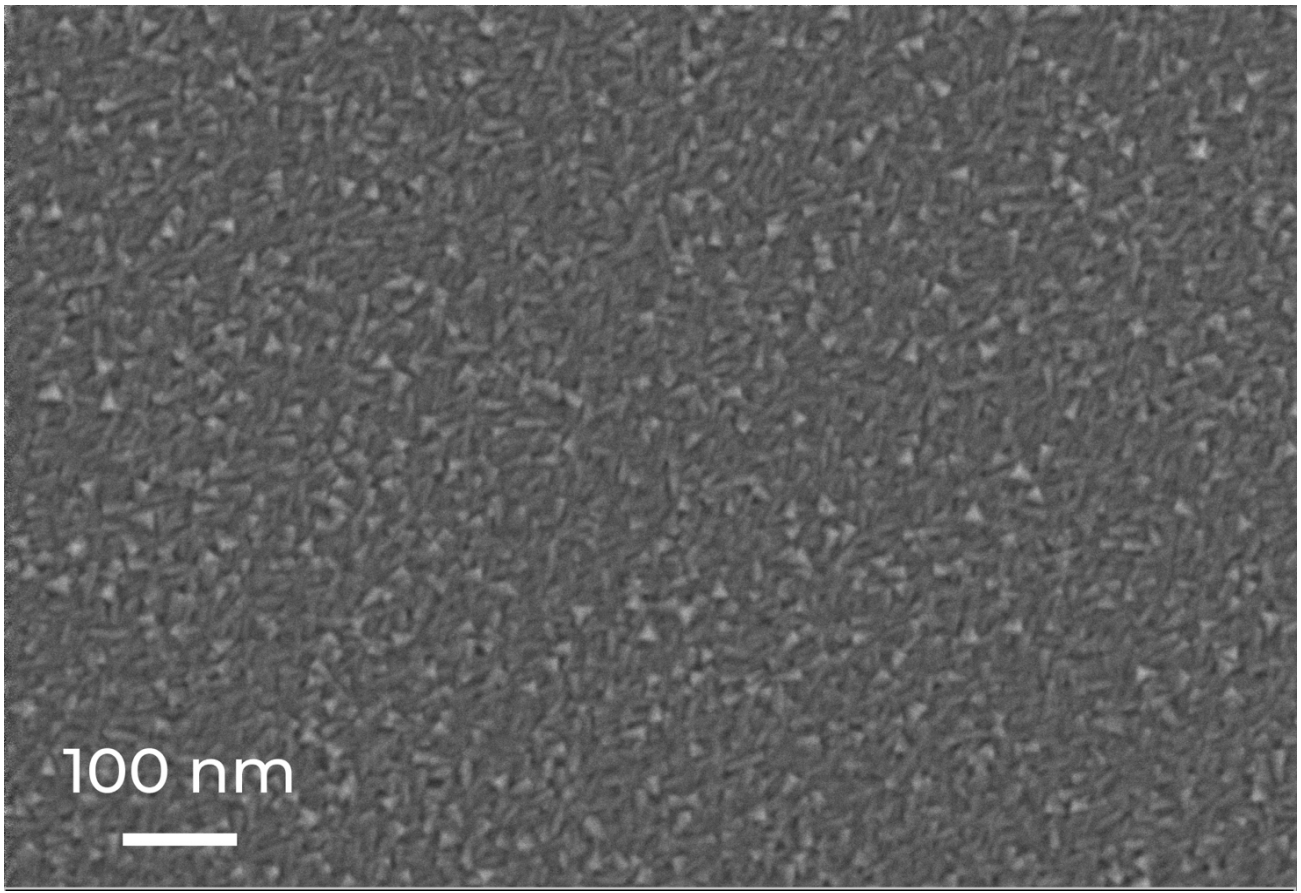

**Figure S11.** SEM image of the surface of an NbN thin film deposited at a temperature of 21 °C and a nitrogen concentration of 20% on a sapphire substrate

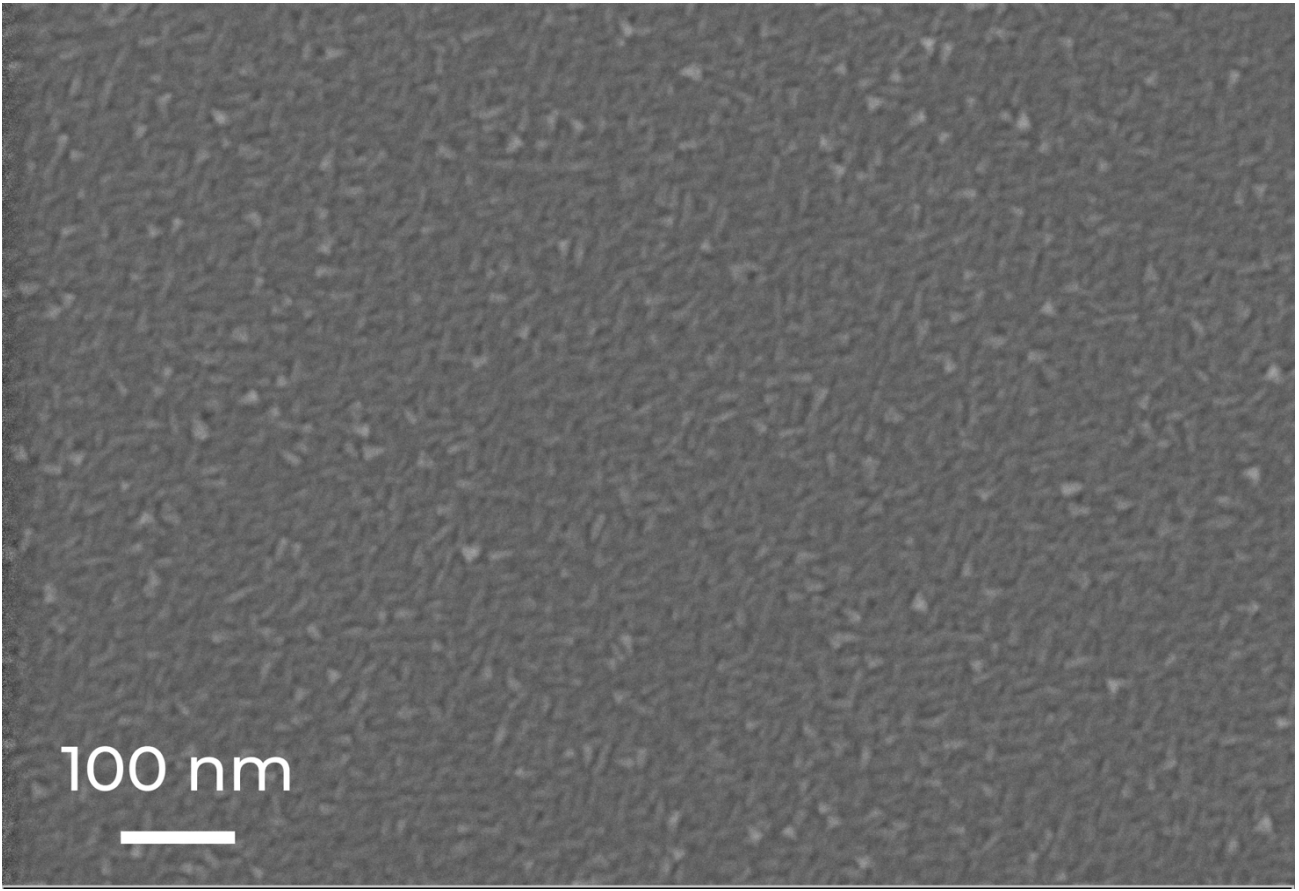

**Figure S12.** SEM image of the surface of an NbN thin film deposited at a temperature of 21 °C and a nitrogen concentration of 35% on a sapphire substrate

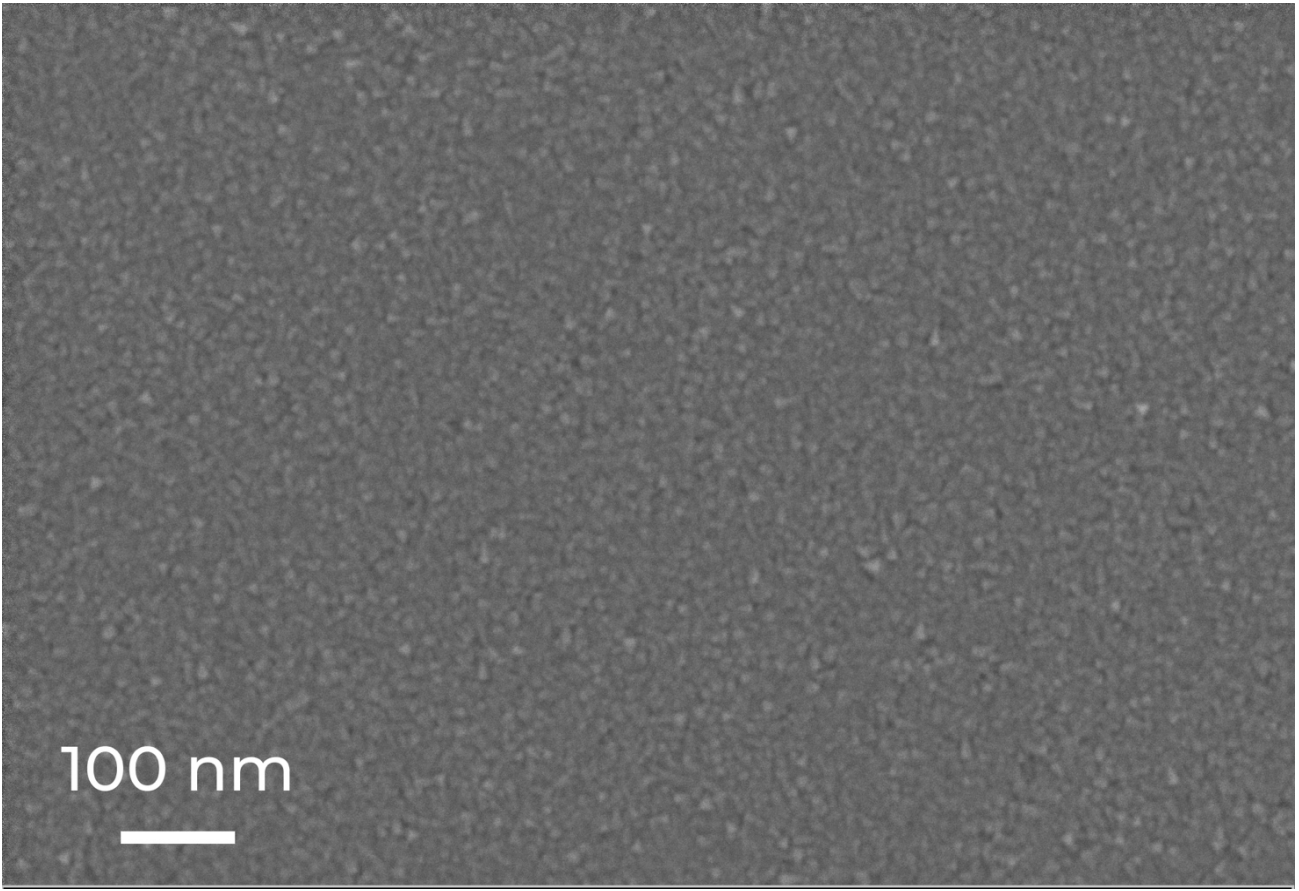

**Figure S13.** SEM image of the surface of an NbN thin film deposited at a temperature of 400 °C and a nitrogen concentration of 10% on a sapphire substrate

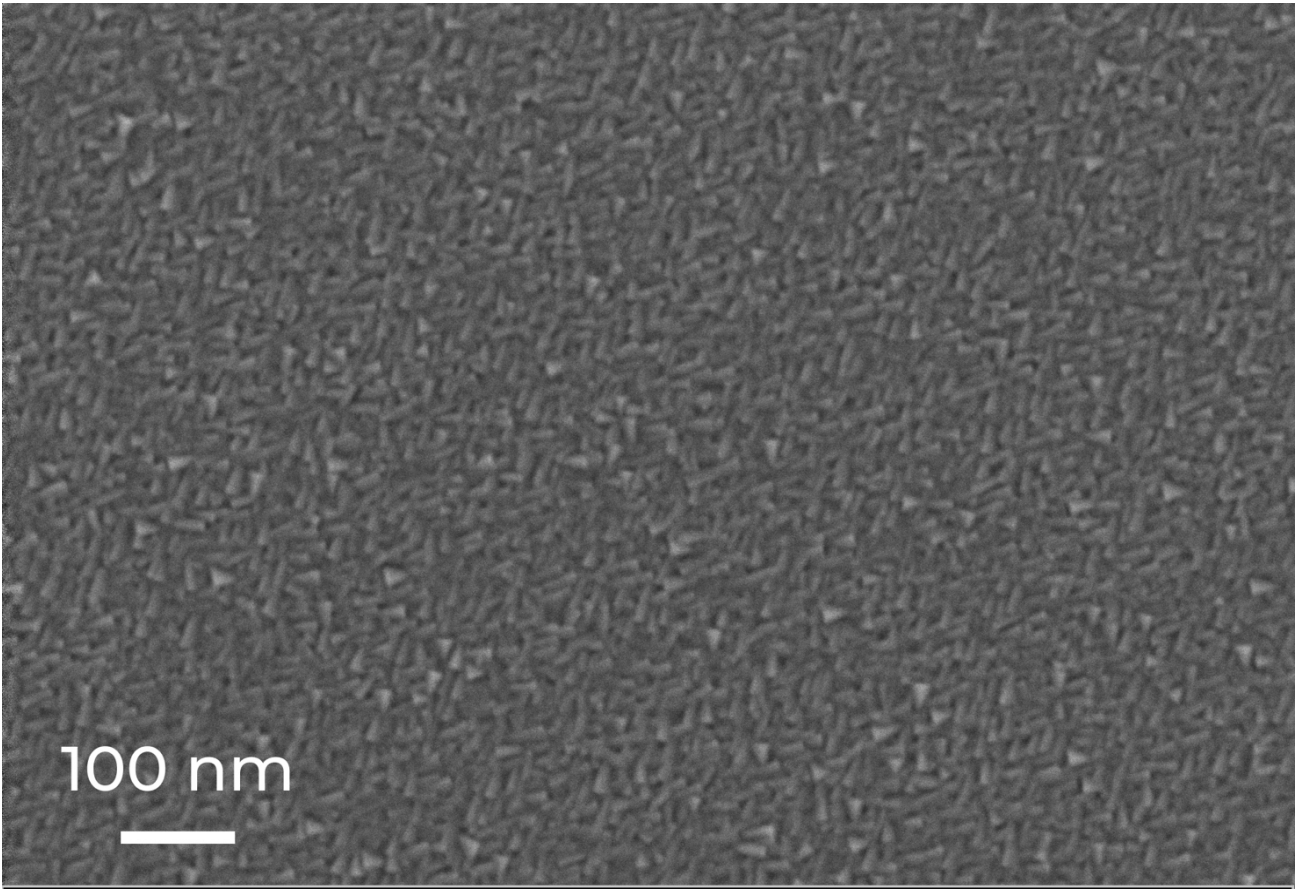

**Figure S14.** SEM image of the surface of an NbN thin film deposited at a temperature of 400 °C and a nitrogen concentration of 20% on a sapphire substrate

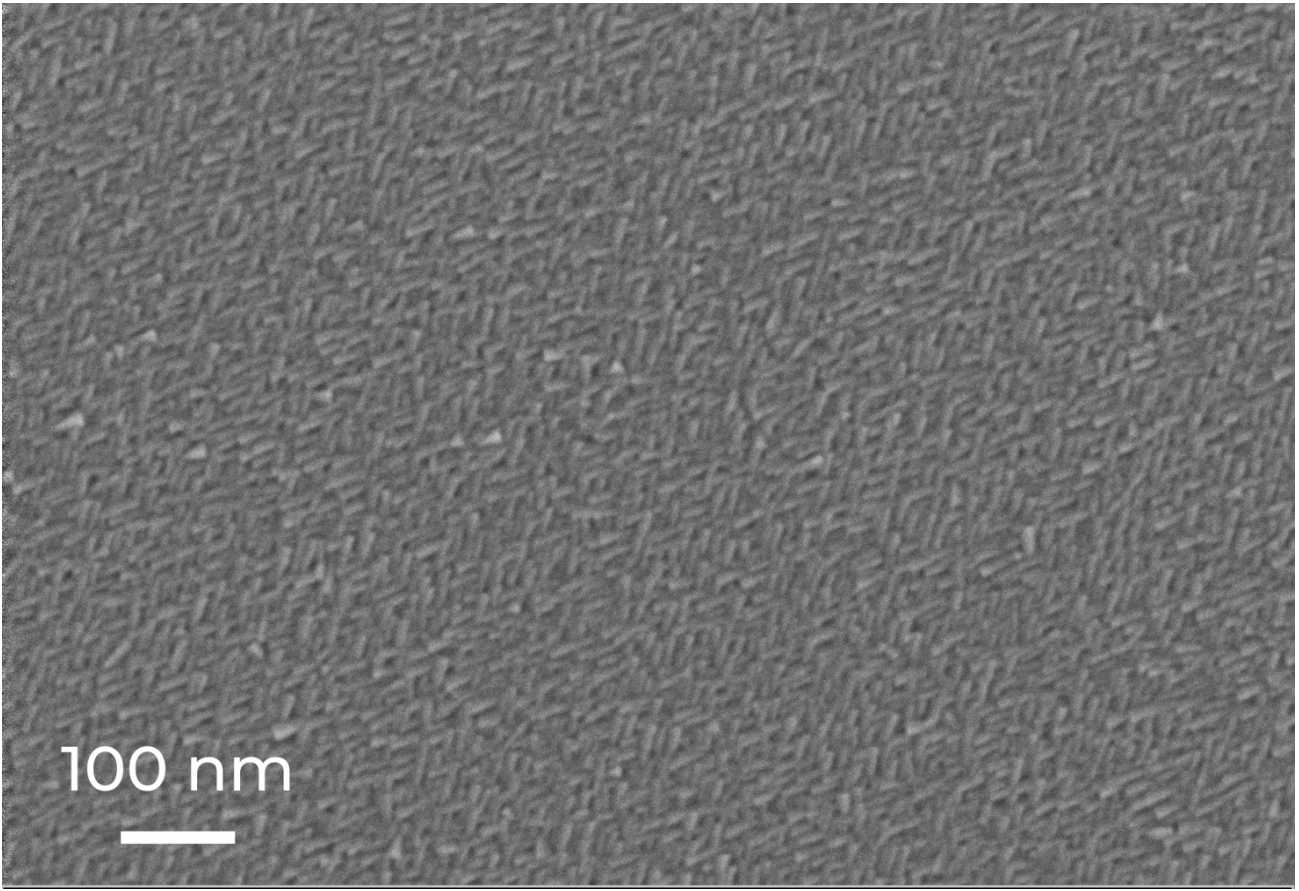

**Figure S15.** SEM image of the surface of an NbN thin film deposited at a temperature of 400 °C and a nitrogen concentration of 35% on a sapphire substrate

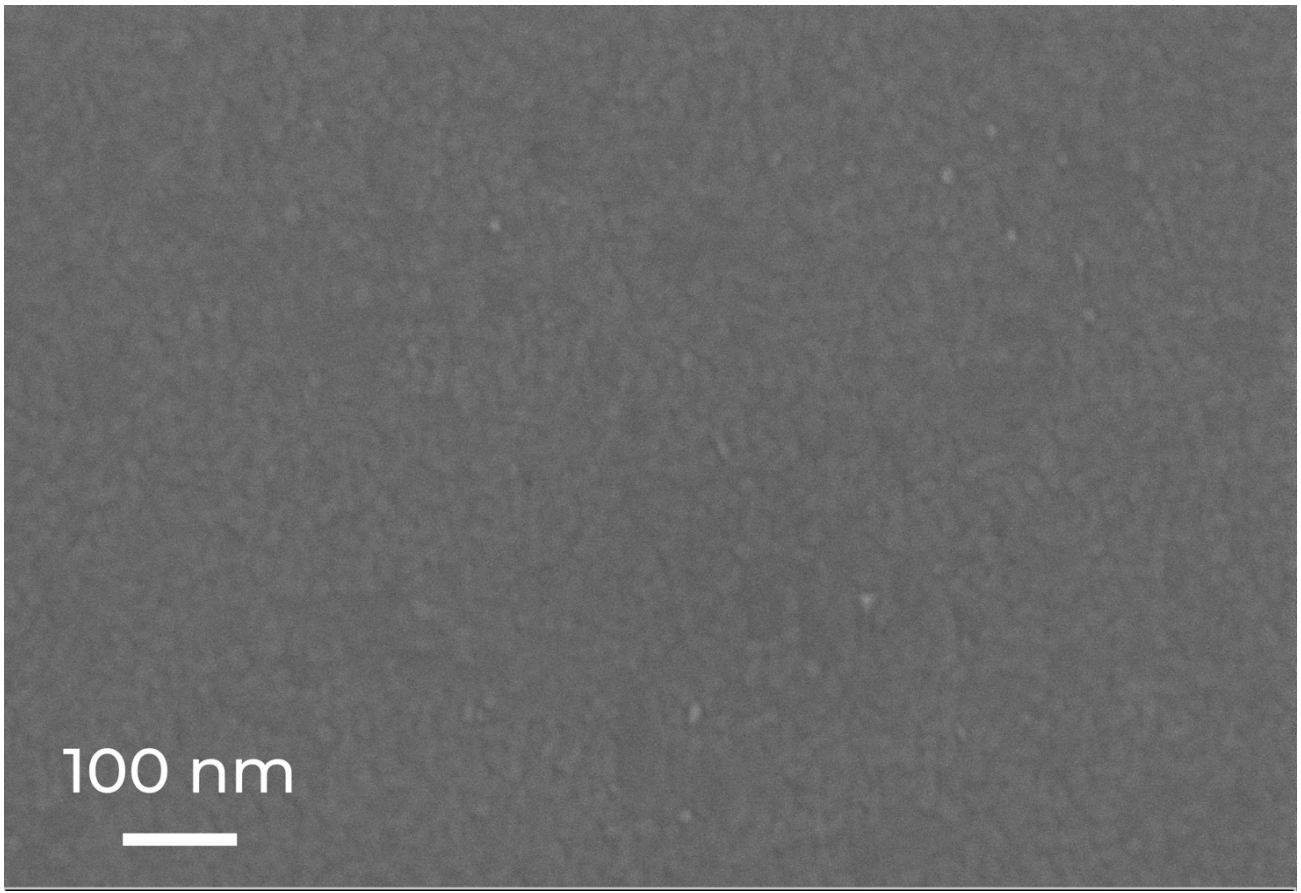

**Figure S16.** SEM image of the surface of an NbN thin film deposited at a temperature of 800 °C and a nitrogen concentration of 10% on a sapphire substrate

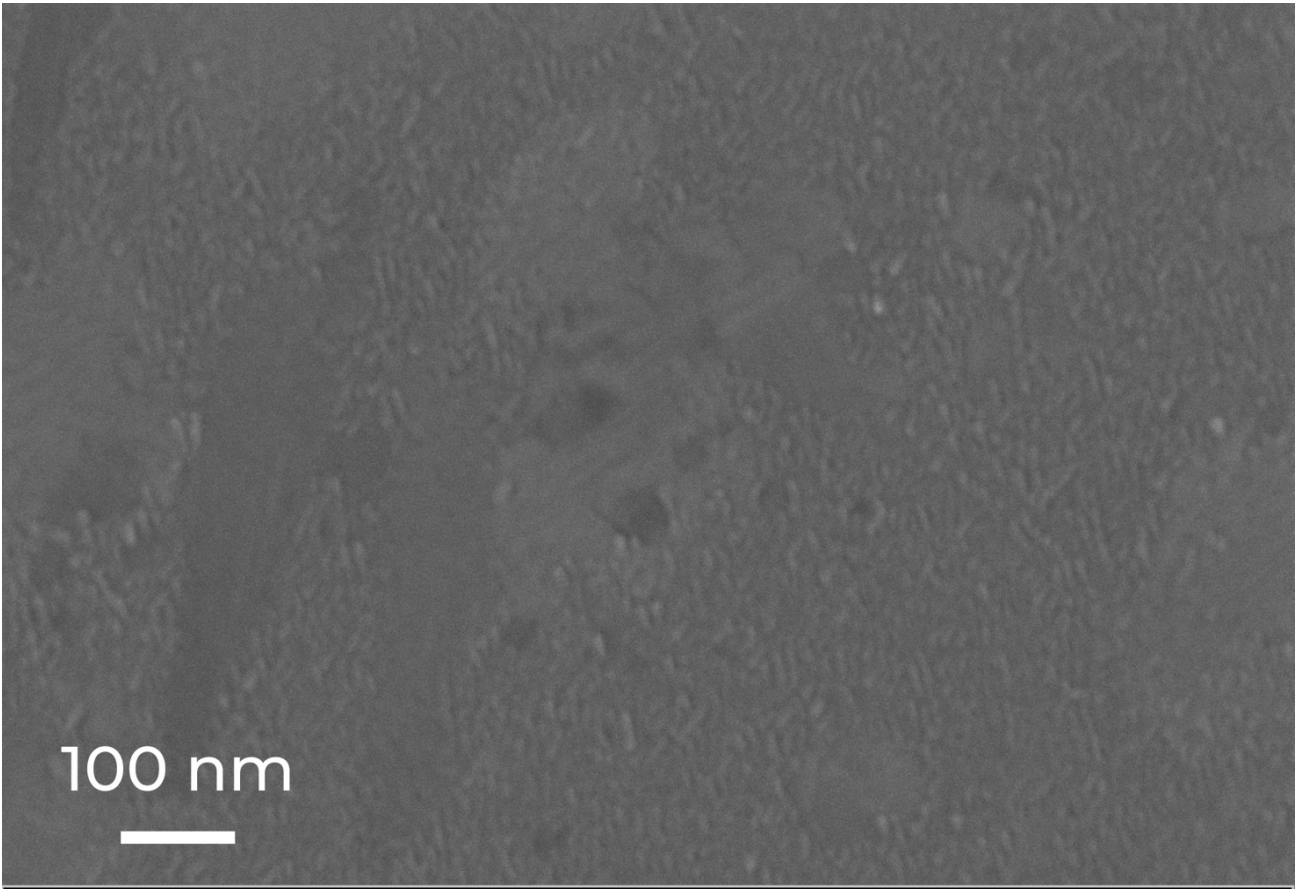

**Figure S17.** SEM image of the surface of an NbN thin film deposited at a temperature of 800 °C and a nitrogen concentration of 20% on a sapphire substrate

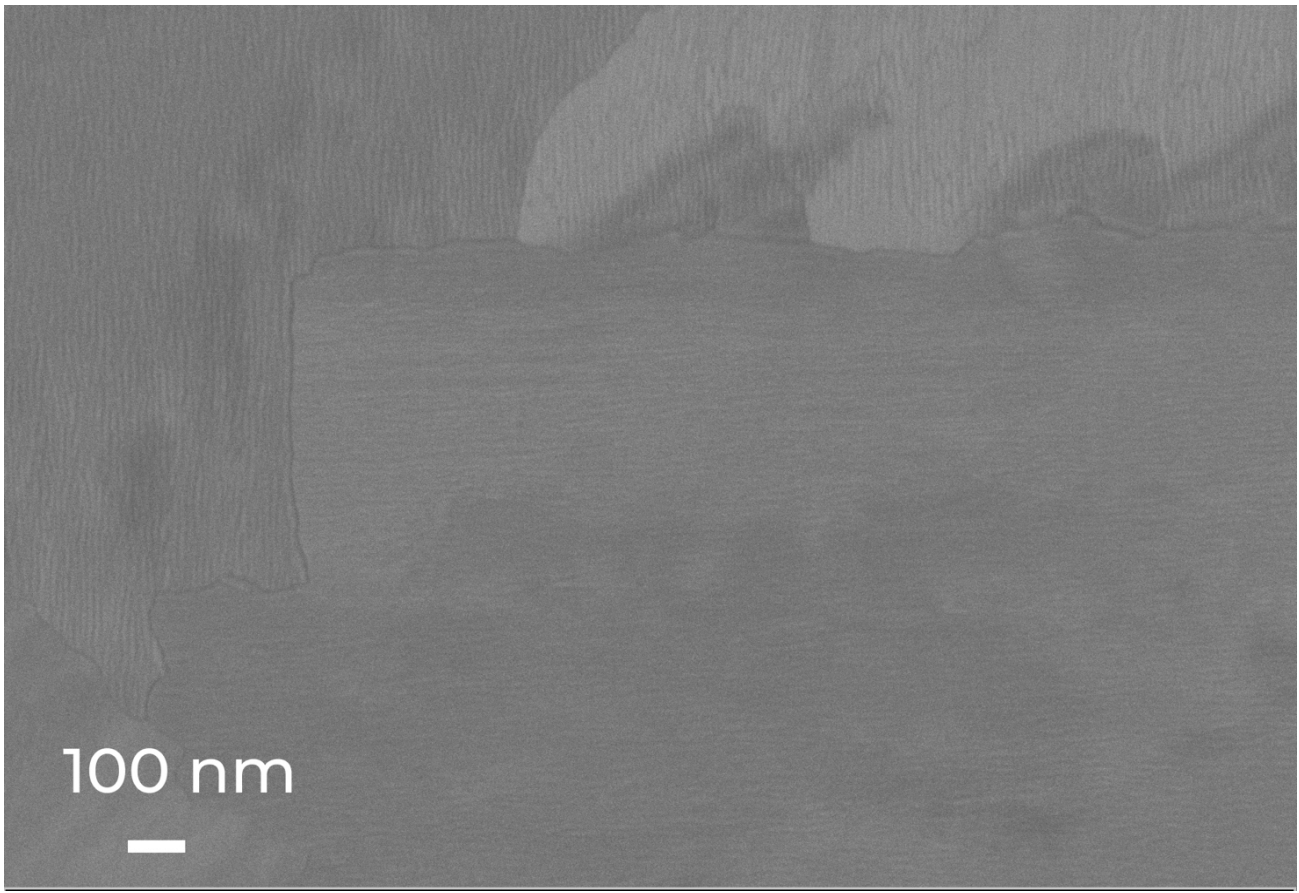

**Figure S18.** SEM image of the surface of an NbN thin film deposited at a temperature of 800 °C and a nitrogen concentration of 35% on a sapphire substrate

**2.3.    NbN thin films on silicon dioxide substrate**

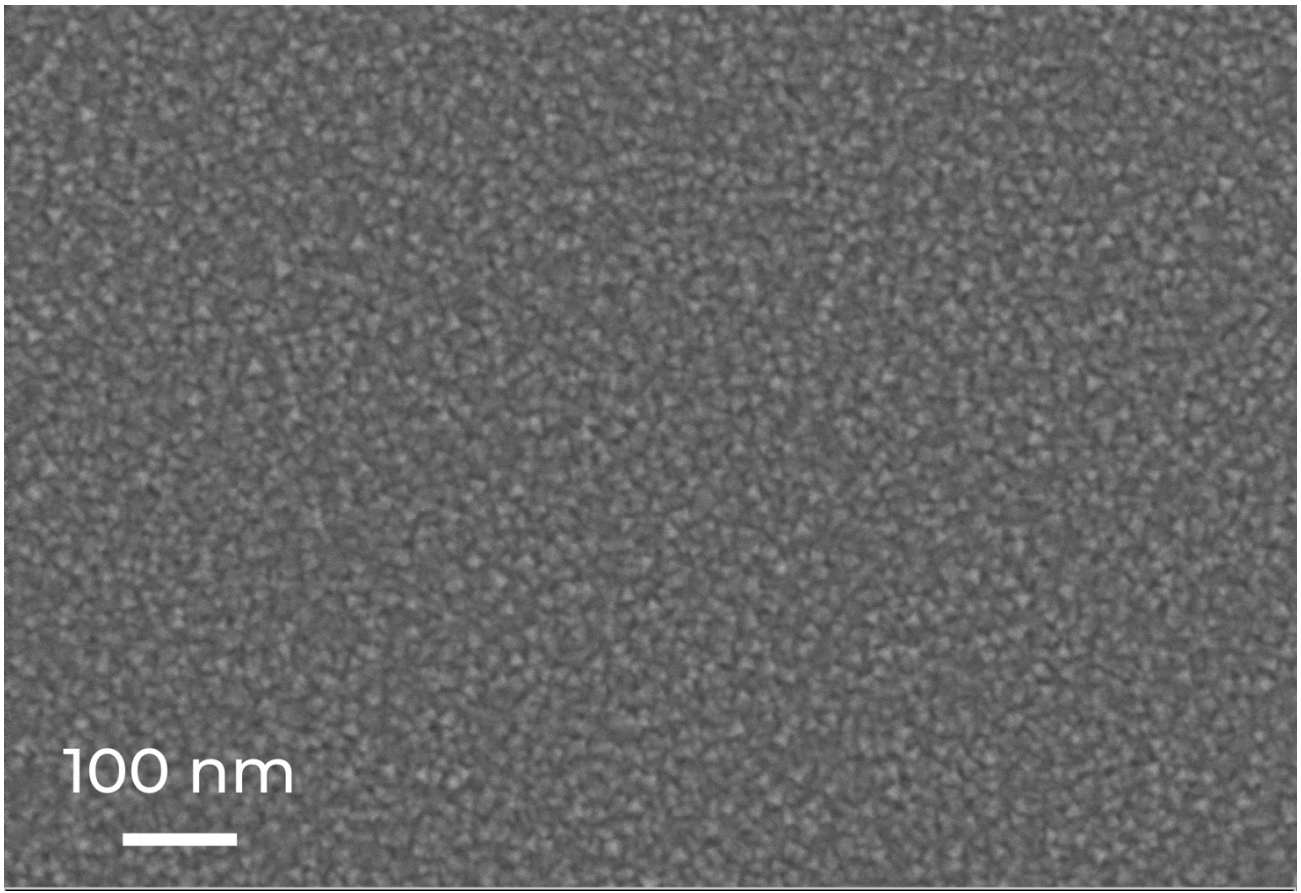

**Figure S19.** SEM image of the surface of an NbN thin film deposited at a temperature of 21 °C and a nitrogen concentration of 10% on a silicon dioxide substrate

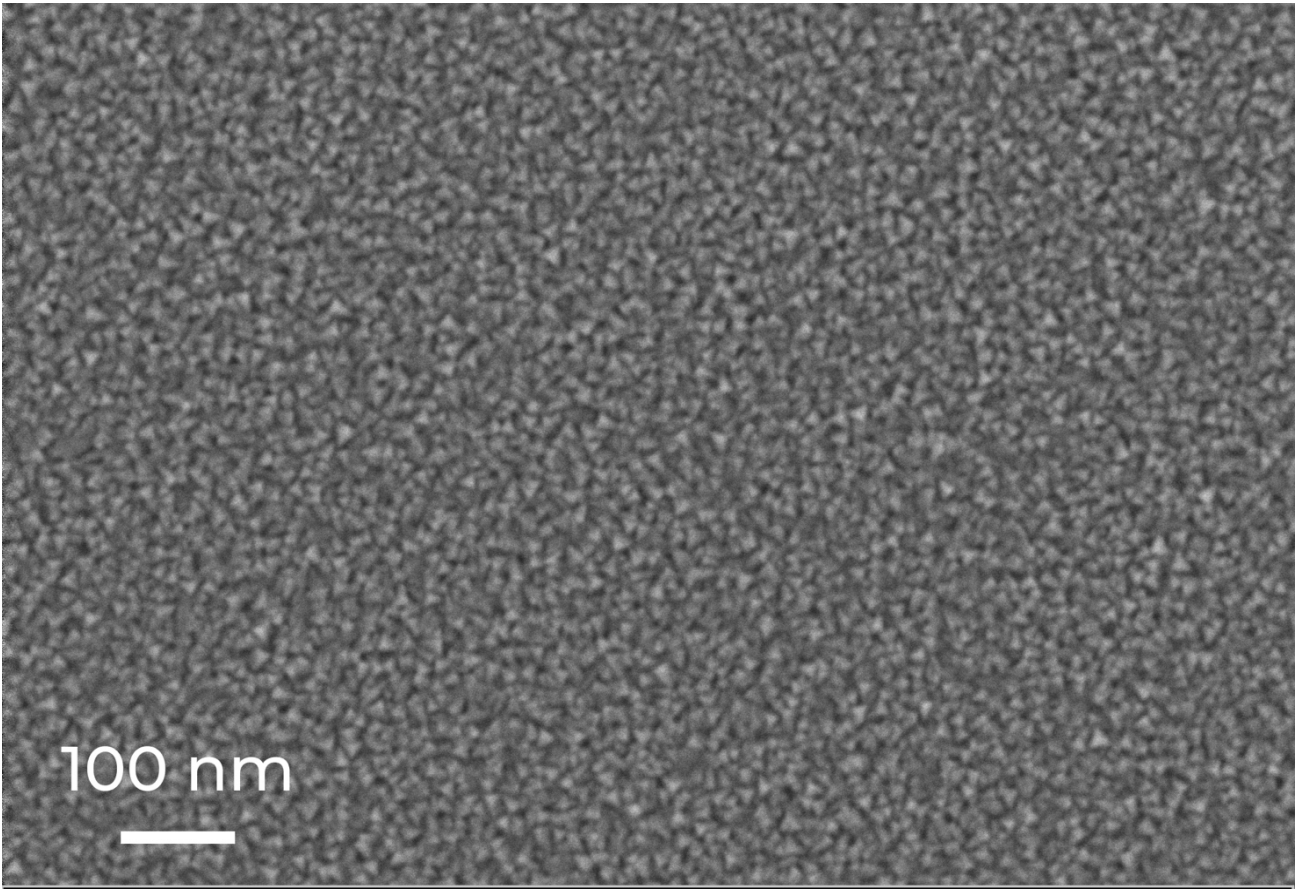

**Figure S20.** SEM image of the surface of an NbN thin film deposited at a temperature of 21 °C and a nitrogen concentration of 20% on a silicon dioxide substrate

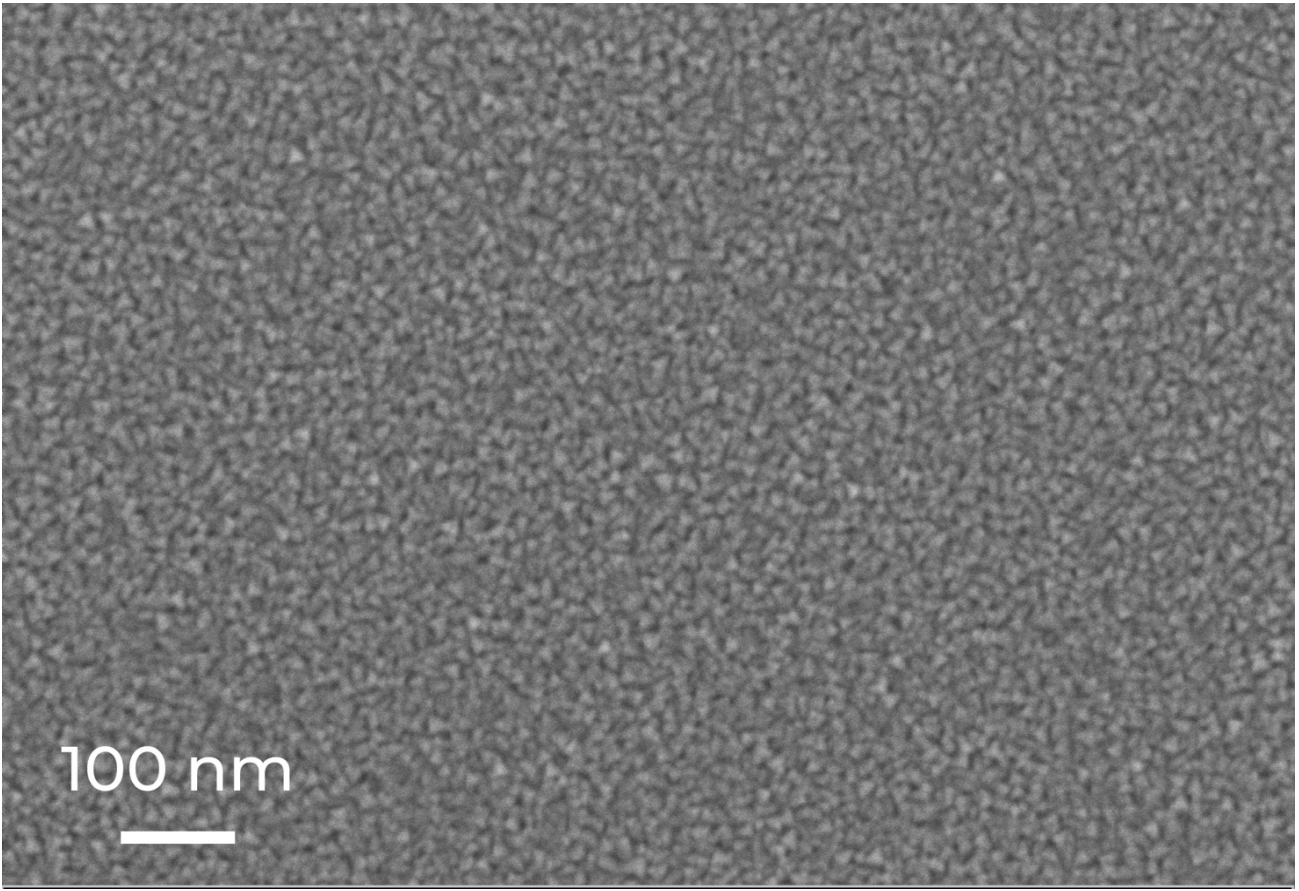

**Figure S21.** SEM image of the surface of an NbN thin film deposited at a temperature of 21 °C and a nitrogen concentration of 35% on a silicon dioxide substrate

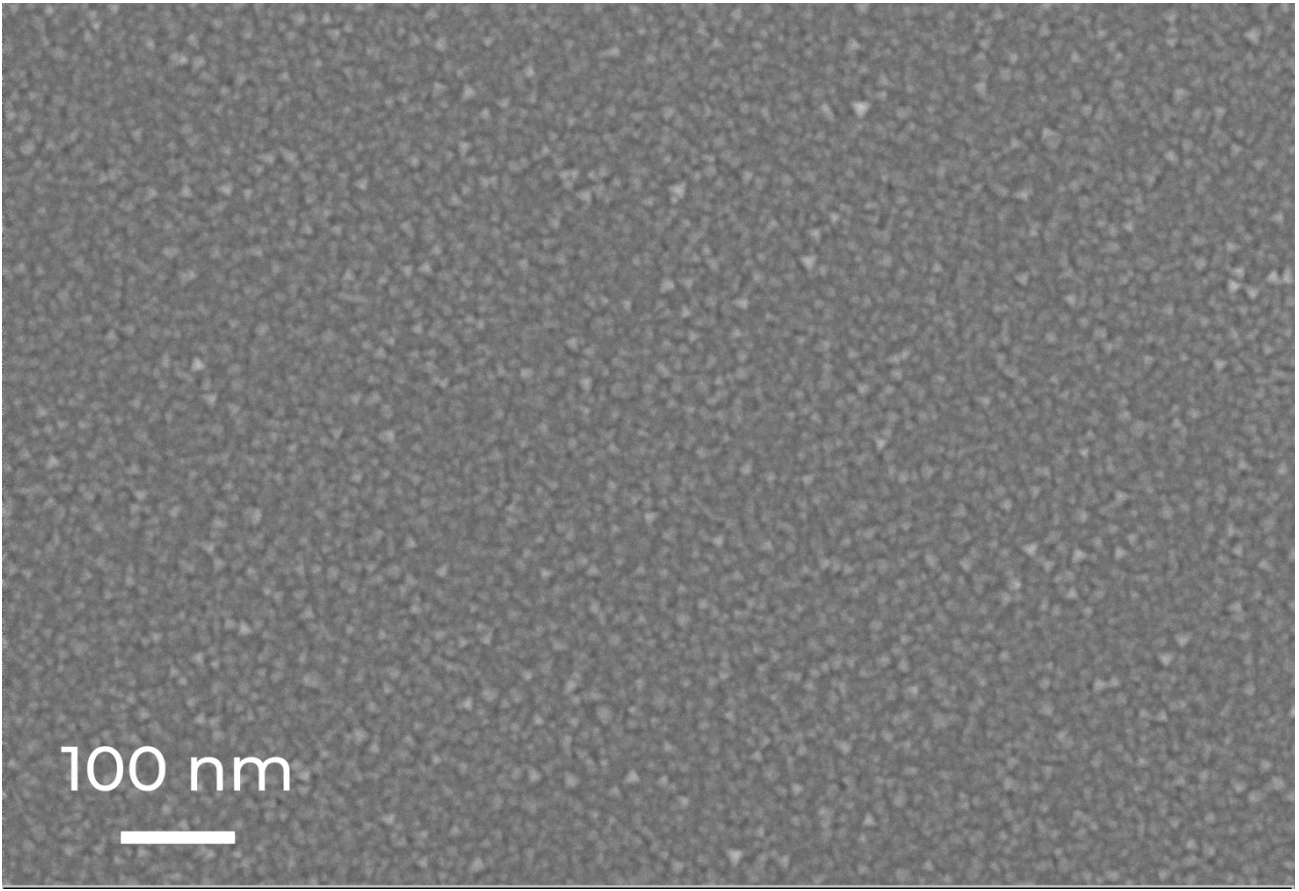

**Figure S22.** SEM image of the surface of an NbN thin film deposited at a temperature of 400 °C and a nitrogen concentration of 10% on a silicon dioxide substrate

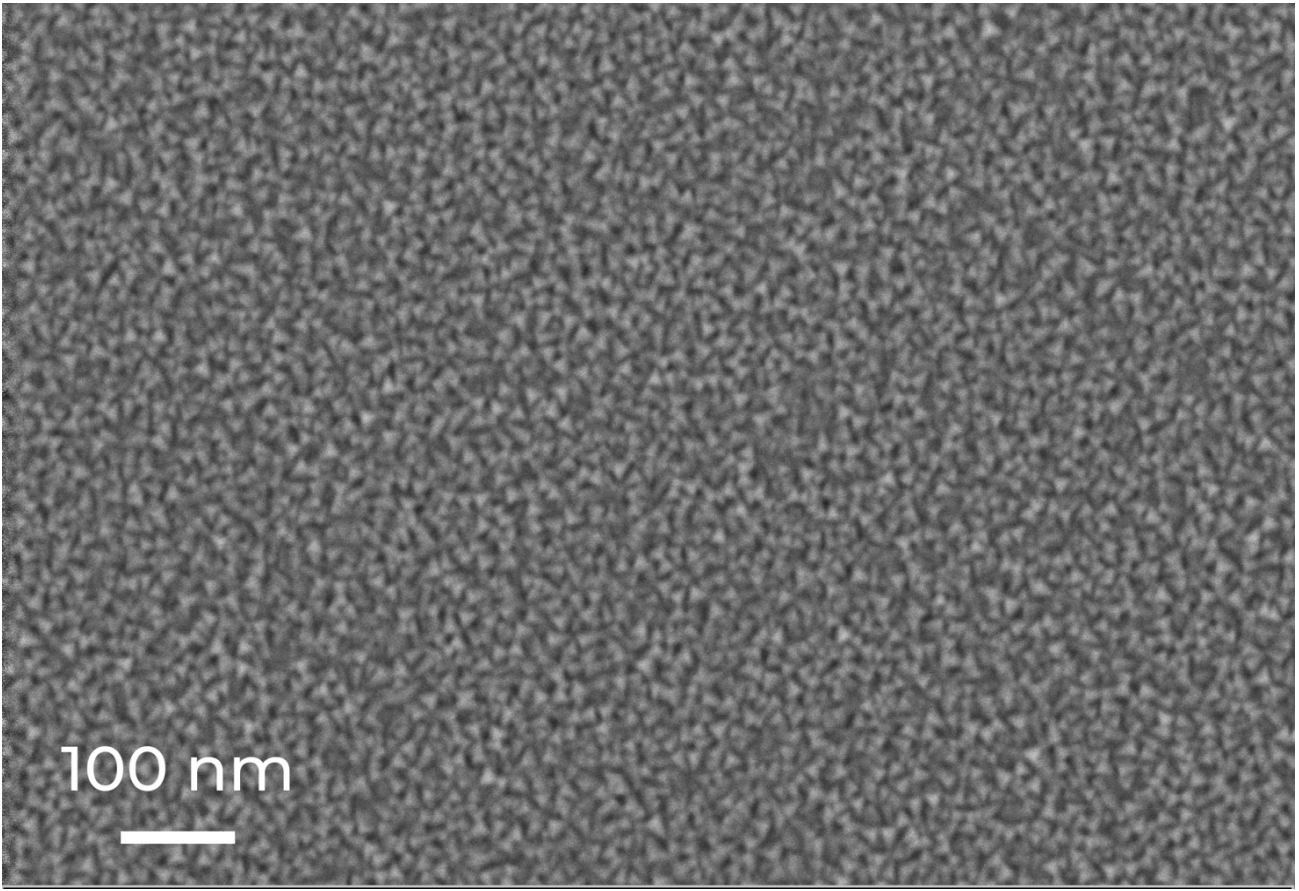

**Figure S23.** SEM image of the surface of an NbN thin film deposited at a temperature of 400 °C and a nitrogen concentration of 20% on a silicon dioxide substrate

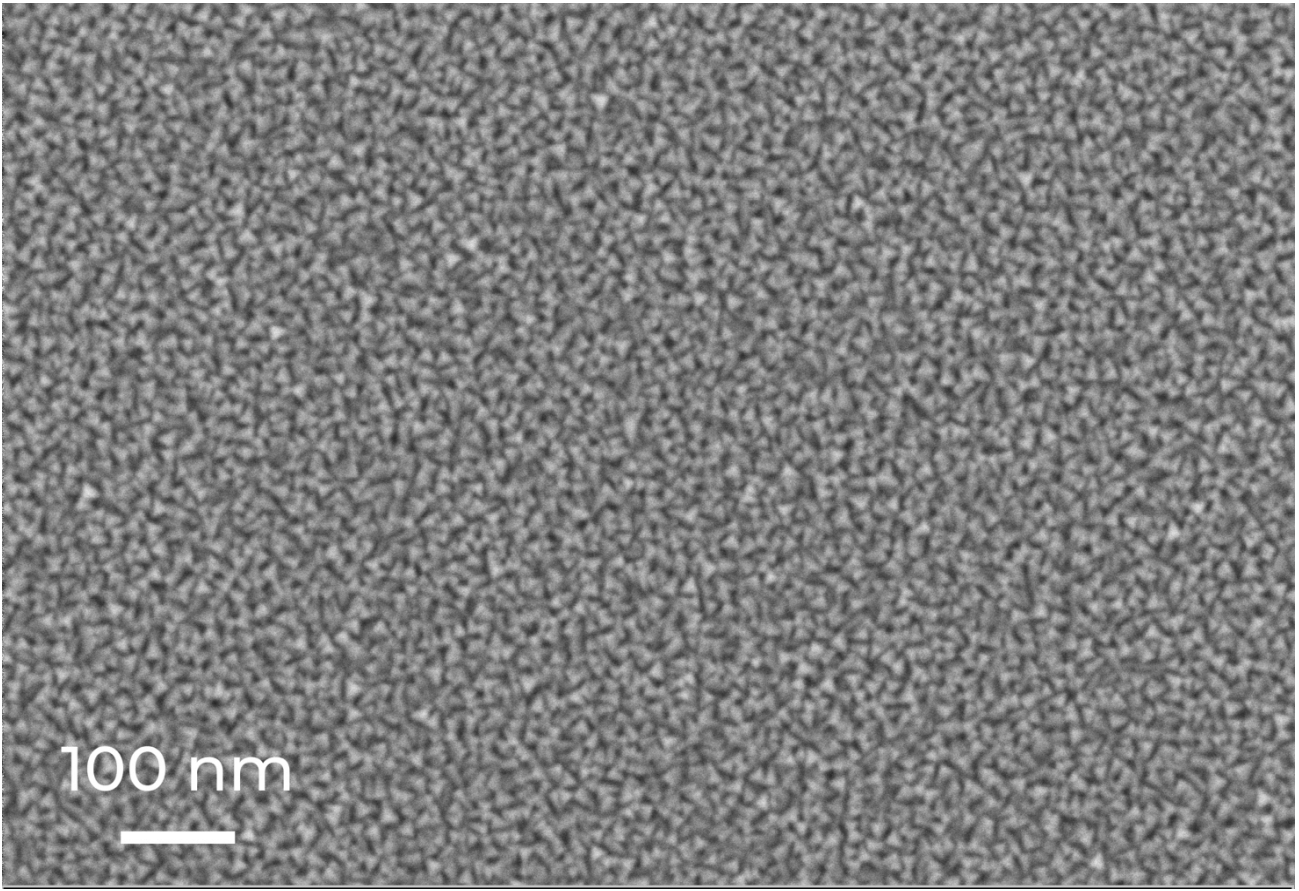

**Figure S24.** SEM image of the surface of an NbN thin film deposited at a temperature of 400 °C and a nitrogen concentration of 35% on a silicon dioxide substrate

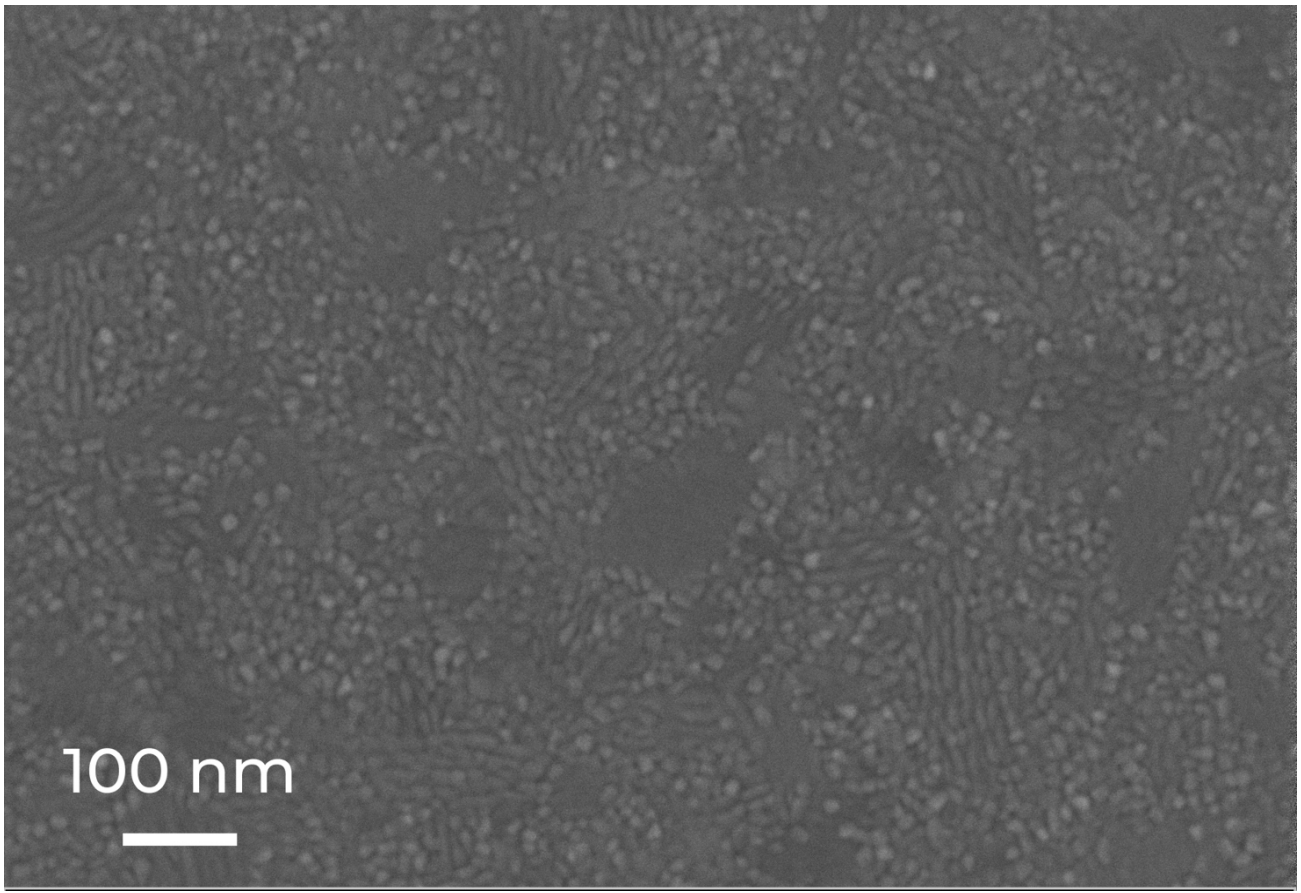

**Figure S25.** SEM image of the surface of an NbN thin film deposited at a temperature of 800 °C and a nitrogen concentration of 10% on a silicon dioxide substrate

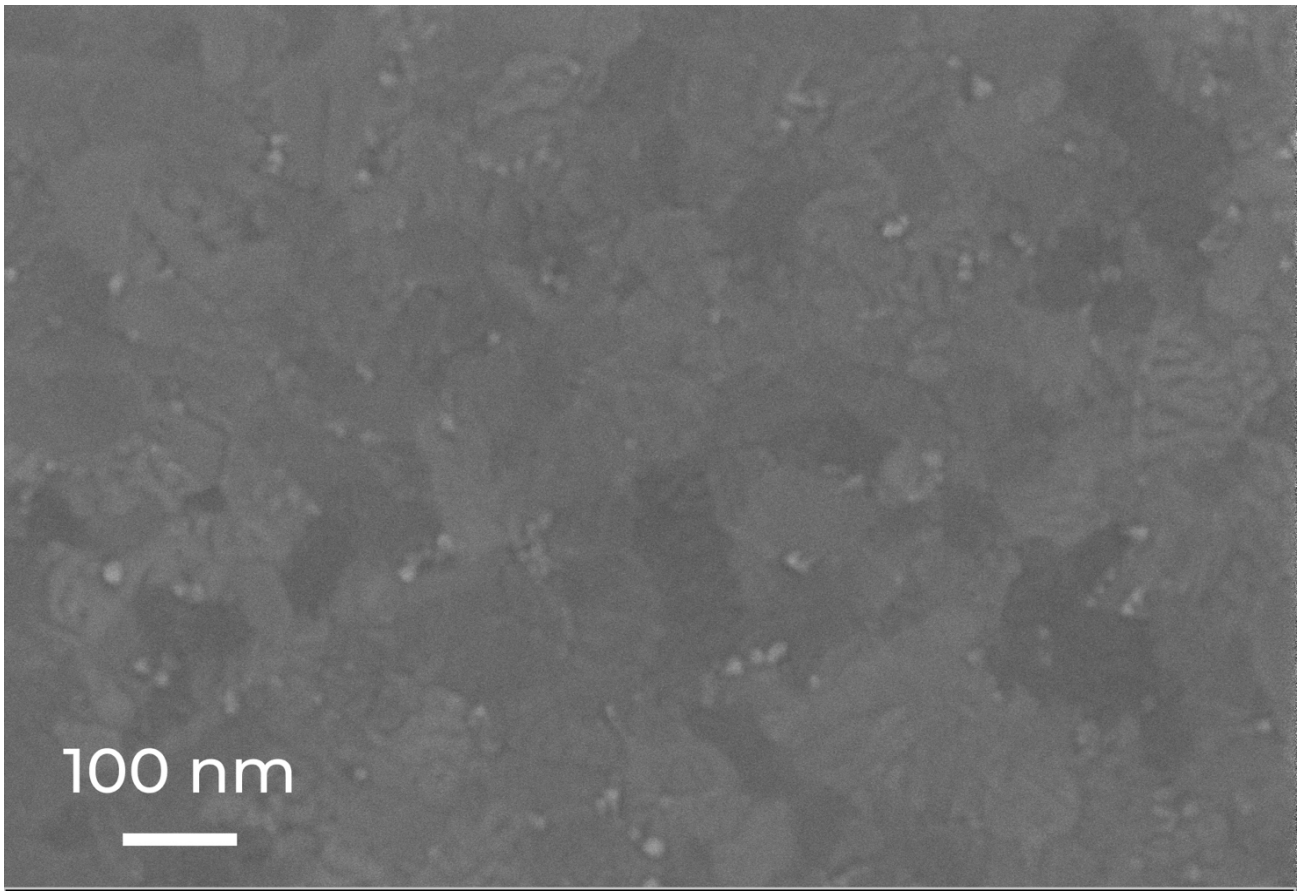

**Figure S26.** SEM image of the surface of an NbN thin film deposited at a temperature of 800 °C and a nitrogen concentration of 20% on a silicon dioxide substrate

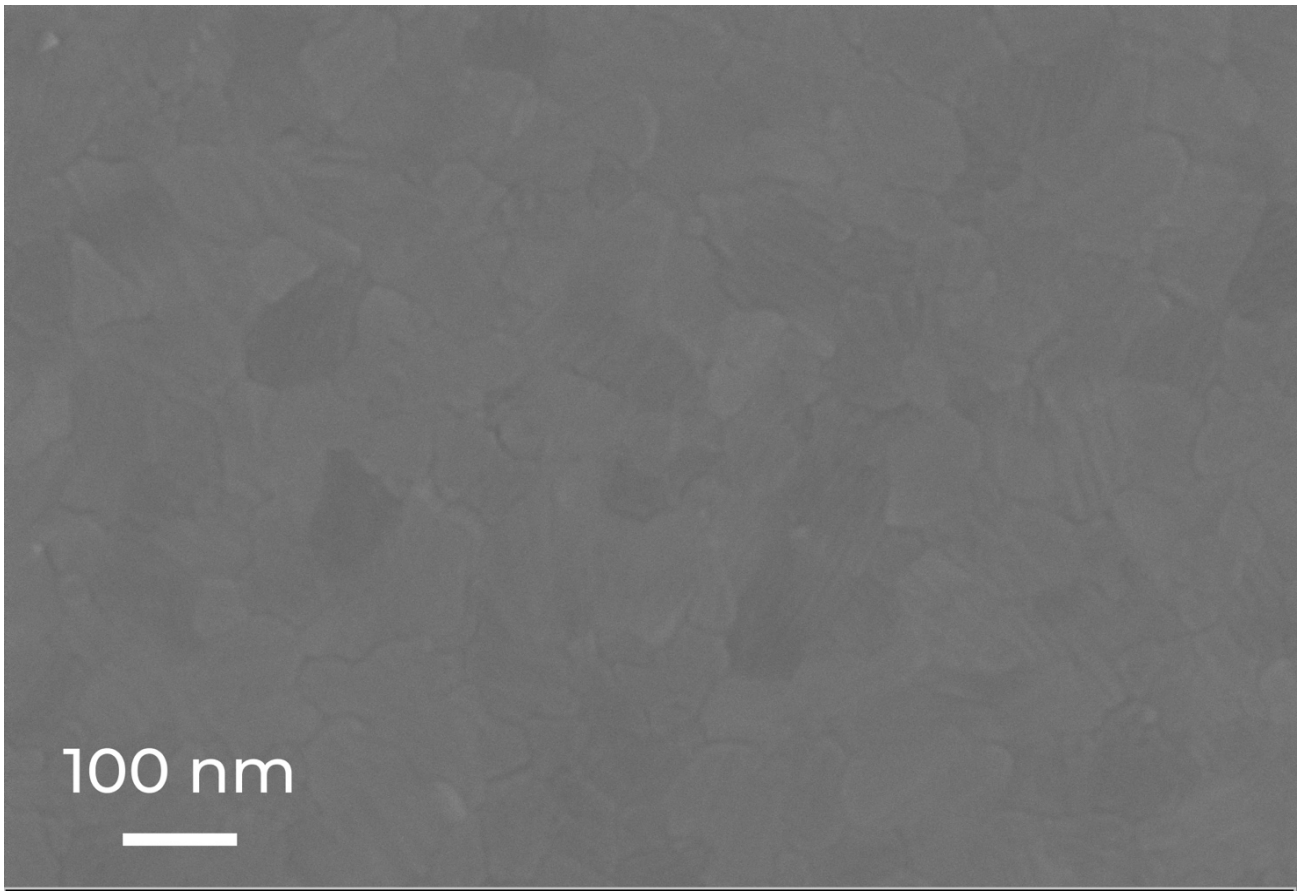

**Figure S27.** SEM image of the surface of an NbN thin film deposited at a temperature of 800 °C and a nitrogen concentration of 35% on a silicon dioxide substrate

**2.4. NbN thin films on silicon nitride substrate**

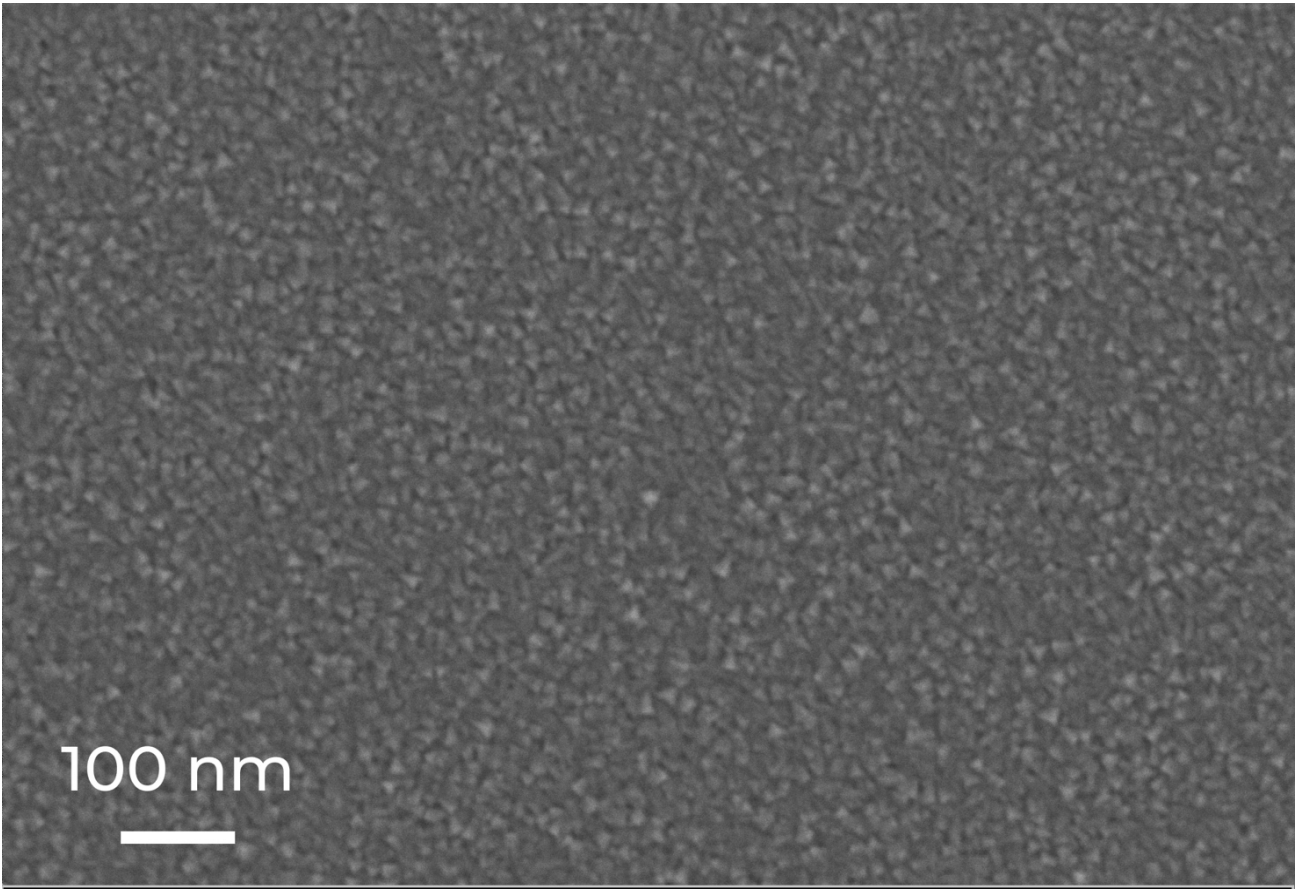

**Figure S28.** SEM image of the surface of an NbN thin film deposited at a temperature of 21 °C and a nitrogen concentration of 10% on a silicon nitride substrate

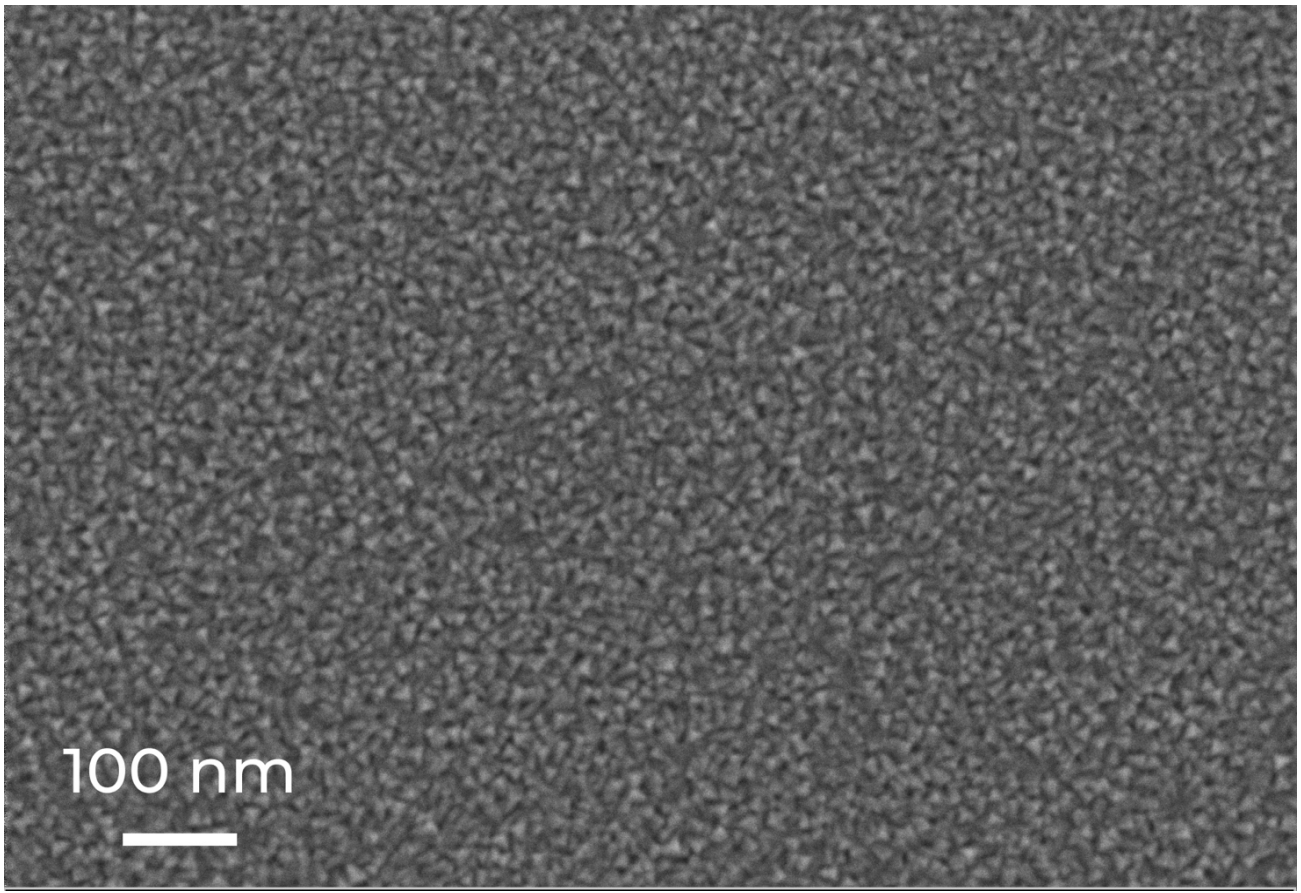

**Figure S29.** SEM image of the surface of an NbN thin film deposited at a temperature of 21 °C and a nitrogen concentration of 20% on a silicon nitride substrate

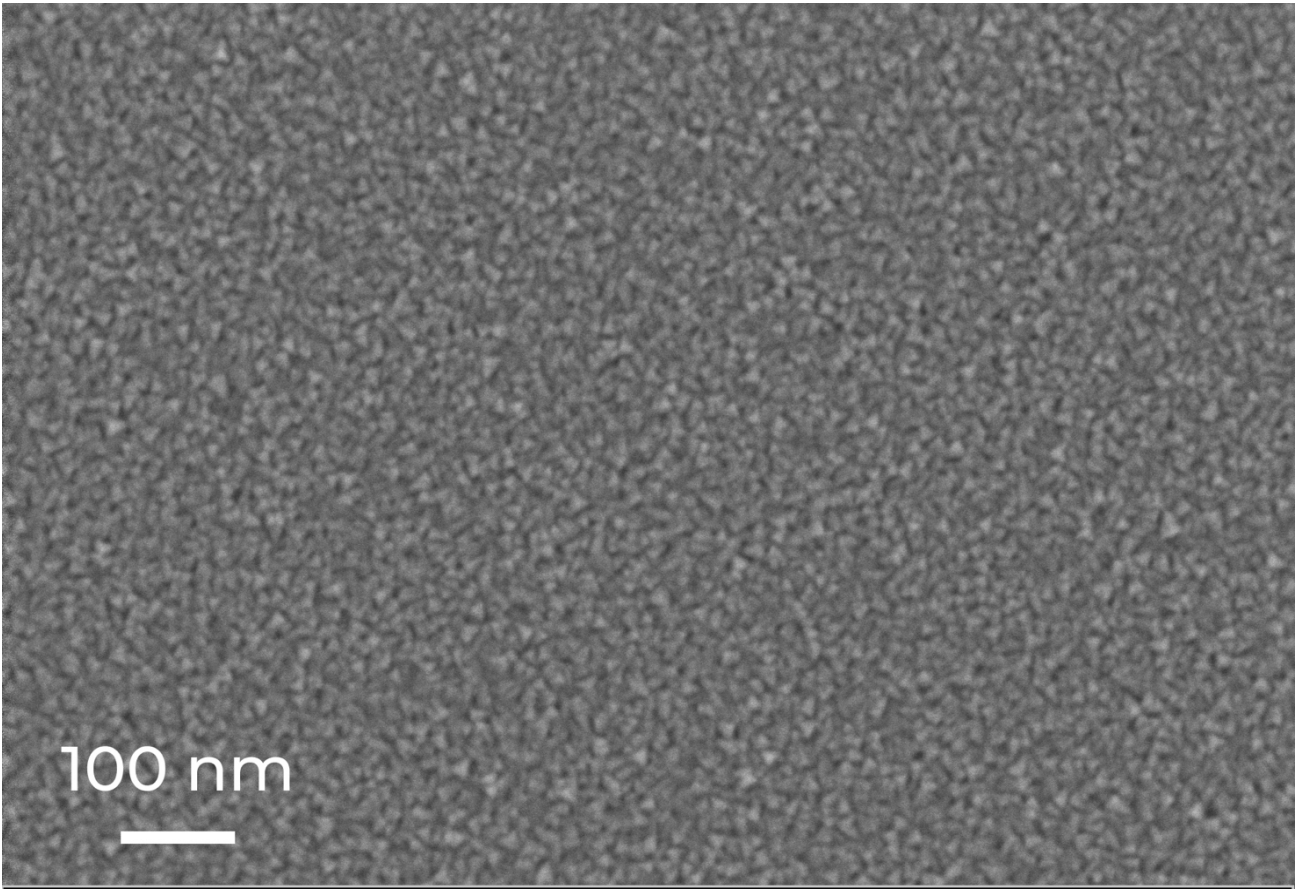

**Figure S30.** SEM image of the surface of an NbN thin film deposited at a temperature of 21 °C and a nitrogen concentration of 35% on a silicon nitride substrate

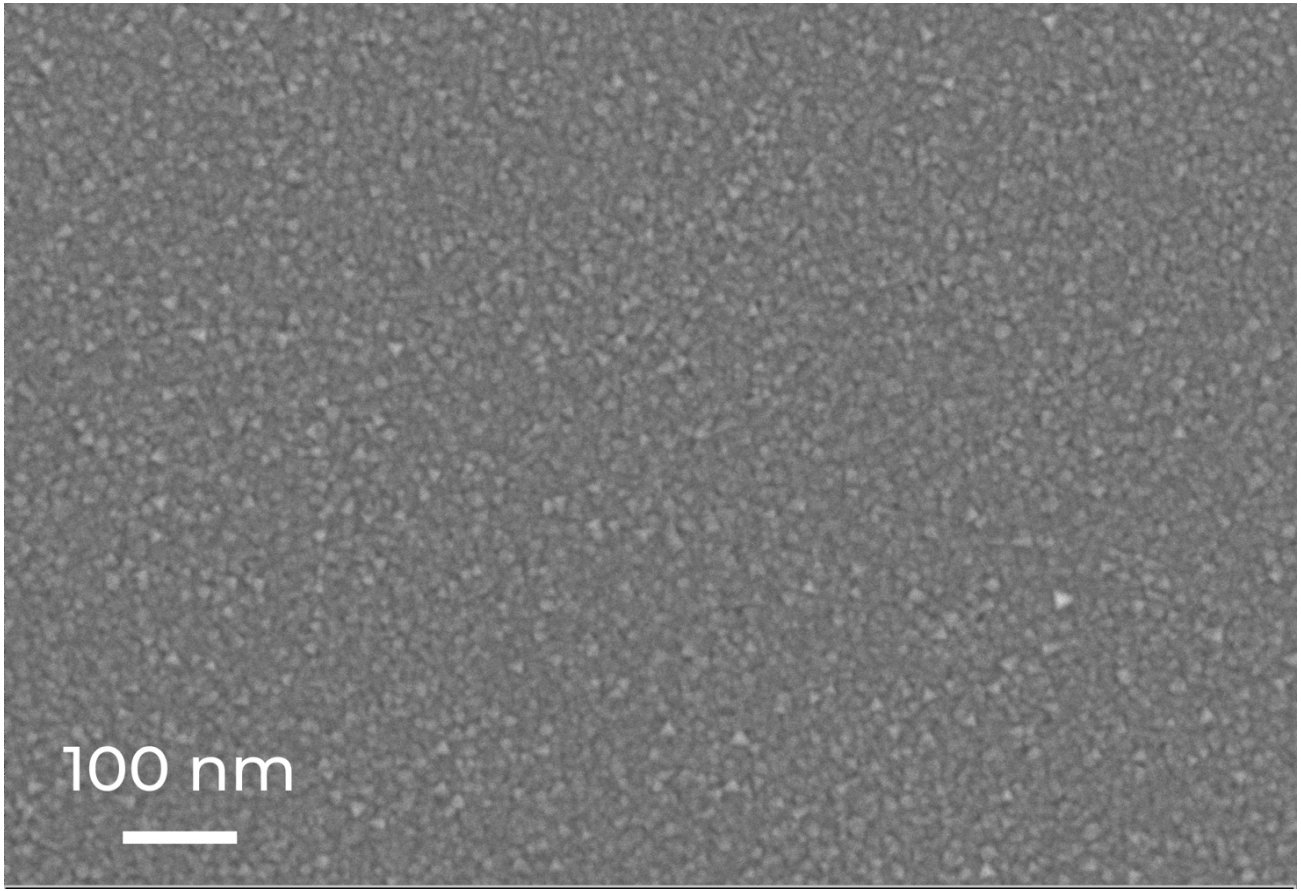

**Figure S31.** SEM image of the surface of an NbN thin film deposited at a temperature of 400 °C and a nitrogen concentration of 10% on a silicon nitride substrate

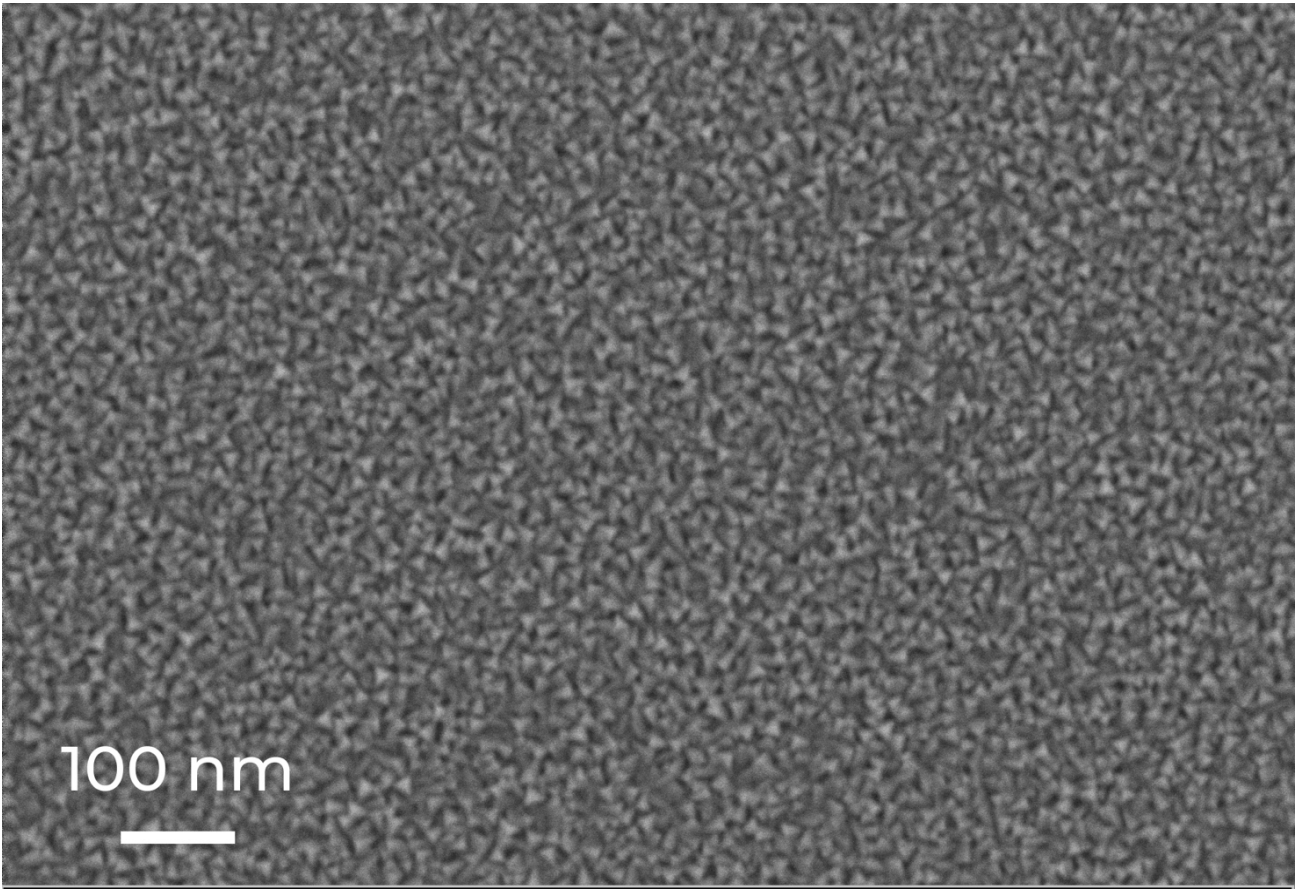

**Figure S32.** SEM image of the surface of an NbN thin film deposited at a temperature of 400 °C and a nitrogen concentration of 20% on a silicon nitride substrate

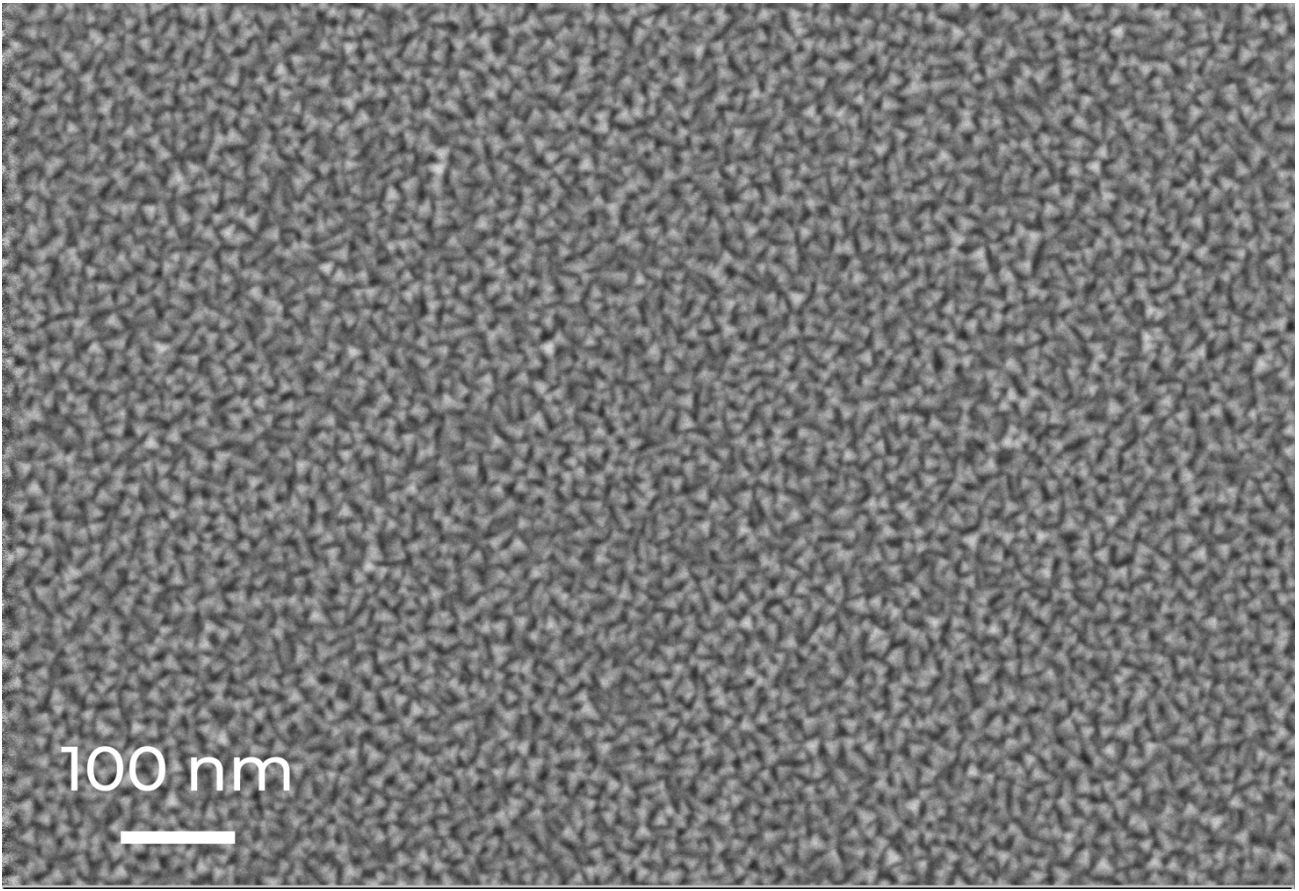

**Figure S33.** SEM image of the surface of an NbN thin film deposited at a temperature of 400 °C and a nitrogen concentration of 35% on a silicon nitride substrate

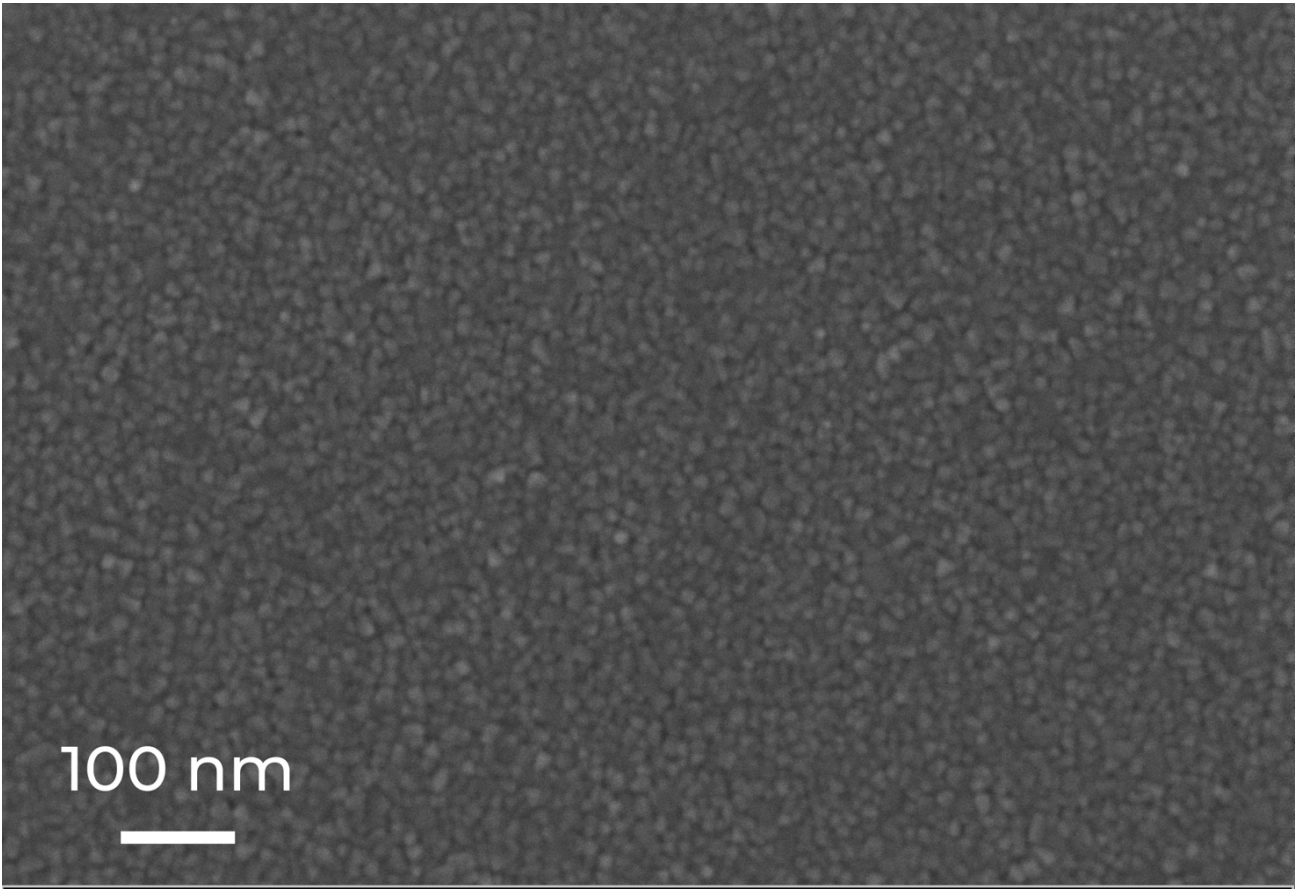

**Figure S34.** SEM image of the surface of an NbN thin film deposited at a temperature of 800 °C and a nitrogen concentration of 10% on a silicon nitride substrate

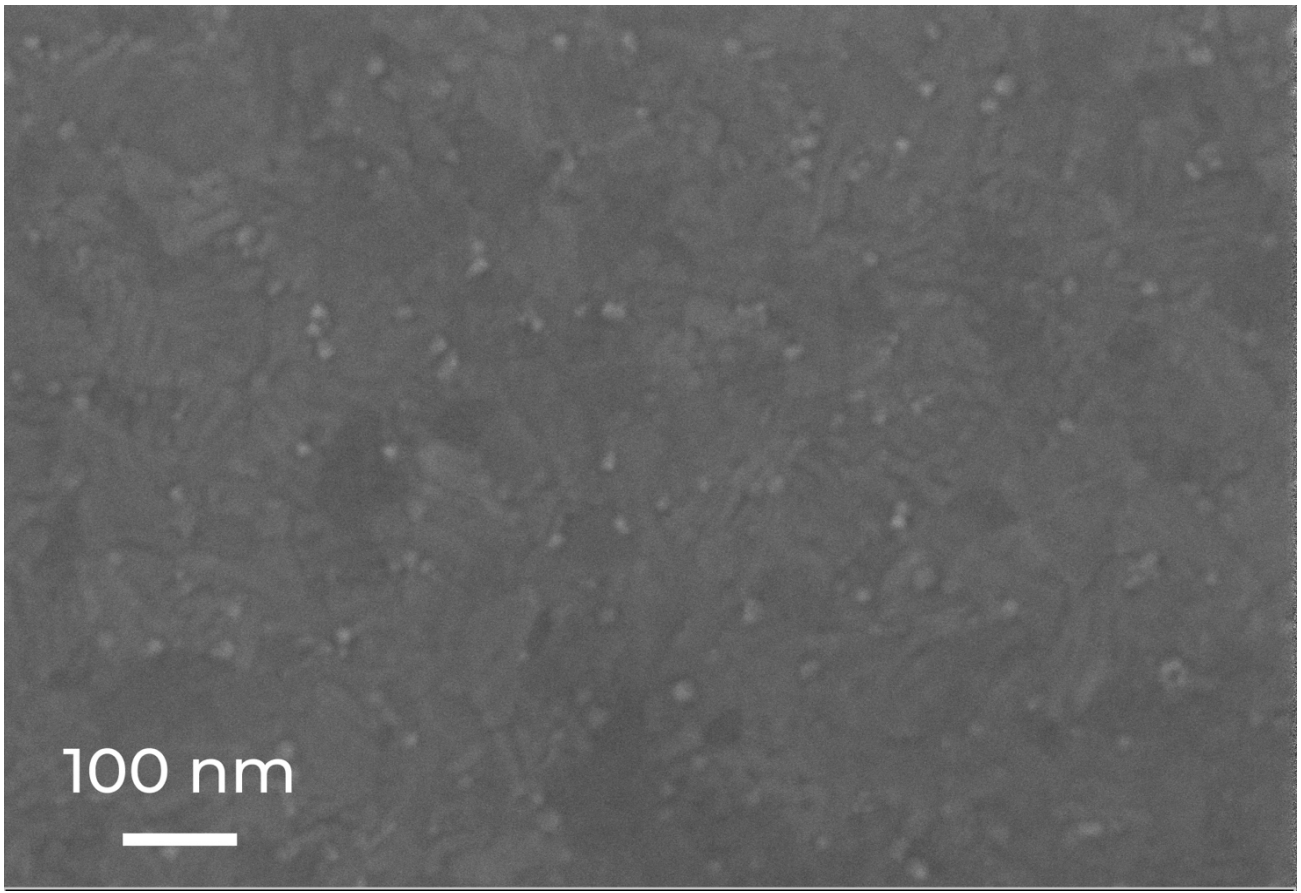

**Figure S35.** SEM image of the surface of an NbN thin film deposited at a temperature of 800 °C and a nitrogen concentration of 20% on a silicon nitride substrate

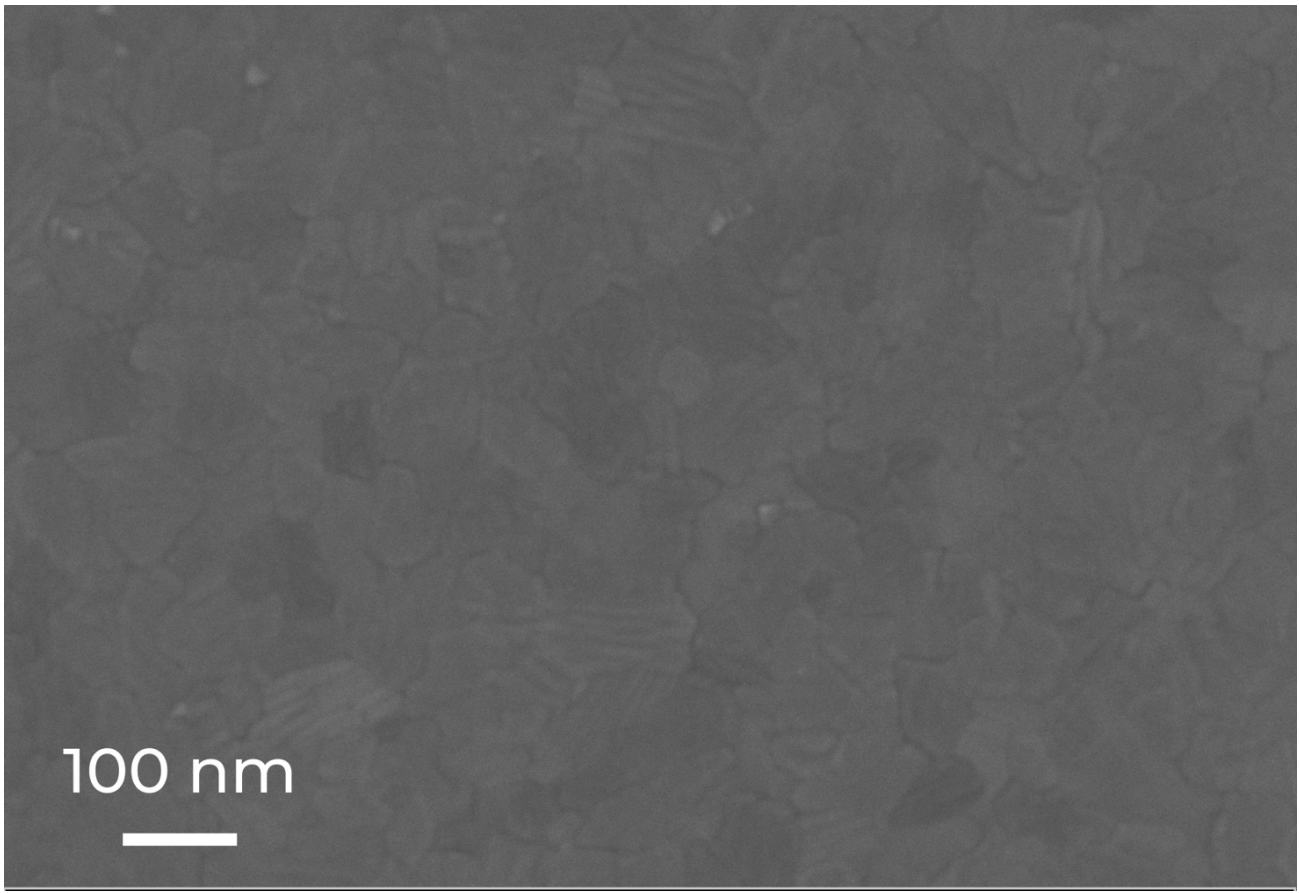

**Figure S36.** SEM image of the surface of an NbN thin film deposited at a temperature of 800 °C and a nitrogen concentration of 35% on a silicon nitride substrate

3. **Ultrathin NbN degradation study**

To study the degradation of NbN film properties over time, we deposited identical ultrathin NbN films on various substrates and monitored their sheet resistances over time. The Figure S37 shows the sheet resistance of NbN films on various substrates depending on the time after exposure to atmosphere after deposition.

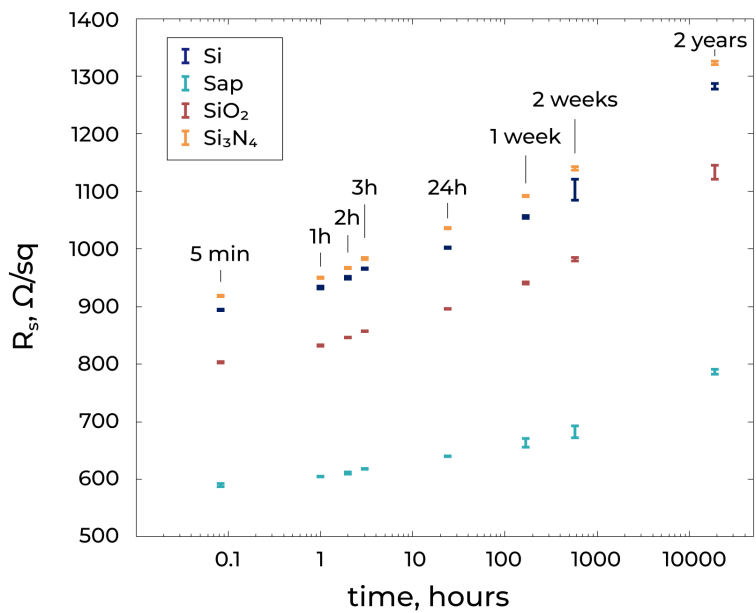

**Figure S37.** Changes in the sheet resistance of NbN films on various substrates with time after exposure to atmosphere.

In order to determine whether the properties of NbN films change due to interaction with the atmosphere or whether the relaxation processes in the film that occur for a certain time after deposition also affect, an additional experiment was carried out. After deposition of an ultrathin NbN film simultaneously on three identical silicon substrates, they were removed from vacuum after different times. After unloading, their sheet resistances were measured immediately. Thus, if relaxation processes in NbN have a significant effect on its properties, then the resistances of films removed from vacuum after different times should differ greatly. Figure S38 shows the sheet resistances of films removed from vacuum after different times. The time of unloading the first film was taken as zero, however, this time was several hours from the moment of the end of deposition, so that the samples were guaranteed to cool down and their temperature did not affect the result.

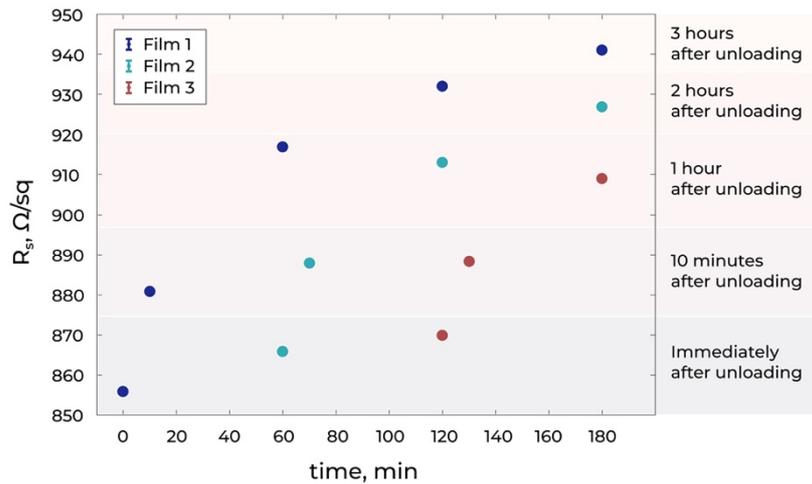

**Figure S38.** Change in sheet resistance with time of three NbN films sputtered in a single process, which were removed from vacuum at various times after the end of deposition.

It can be seen that the time taken to unload the film from the vacuum does not provide a permanent increase or decrease in the $R_s$. A slight increase in sheet resistance immediately after unloading for the second and third films compared to the first one can be explained by the fact that during the unloading of each film, when there was atmospheric pressure for a short time in the load lock, all films interacted with the atmosphere and slightly changed their properties.

## 4. Ultrathin NbN degradation study

### 4.1. NbN deposition

NbN films were deposited on high-resistivity silicon substrates (> 10000 $\Omega$cm), R-plane sapphire, oxidized silicon with a $SiO_2$ thickness of 3.5 $\mu$m and oxidized silicon with 150 nm thick silicon nitride on top. Before deposition, we cleaned the substrates in a Piranha solution to remove organic contaminants, and then, for silicon substrates, the natural oxide was removed in 2% hydrofluoric acid solution. Deposition was carried out using DC reactive magnetron sputtering at a base pressure below $10^{-8}$ Torr. An Nb target with a purity of 99.95% was sputtered in a mixture of argon and nitrogen, both with a purity of 99.9999%.

### 4.2. Sheet resistance measurement

The sheet resistance was measured using a four-probe method using a four-probe measurement system. For each test sample, measurements were carried out at least at five points.

### 4.3. Critical temperature measurement

Critical temperature measurements were carried out in a closed-cycle delusion cryostat using the four-probe method. A current significantly less than the critical current of the film was supplied by two probes, and the voltage was read out from the other two probes. In the temperature range near the critical temperature, the dependence of the film resistance on temperature was plotted, from which the value of the critical temperature was determined.

**Scanning electron microscopy**

All deposited NbN thin films were examined using a scanning electron microscope. All SEM images were obtained using an in-lens detector and an accelerating voltage of 5 kV.

**Supplementary Information References**